\documentclass[aip,jcp,nofootinbib,nobibnotes,amsmath,amssymb,amsfonts,noshowpacs,floatfix]{revtex4}%aps

\usepackage{graphicx,bm,epsfig,setspace}
\usepackage{color}
\usepackage{subfigure}
\usepackage{ulem}
\usepackage{soul}
\usepackage{mathtools}
\makeatletter
\def\graphicscale{\twocolumn@sw{0.33}{0.4}}
\makeatother

\def\spose#1{\hbox to 0pt{#1\hss}}
\def\lesssim{\mathrel{\spose{\lower 3pt\hbox{$\mathchar"218$}}
 \raise 2.0pt\hbox{$\mathchar"13C$}}}
\def\gtrsim{\mathrel{\spose{\lower 3pt\hbox{$\mathchar"218$}}
 \raise 2.0pt\hbox{$\mathchar"13E$}}}

\def\<{\langle}
\def\>{\rangle}
 \def\({\left(}
\def\){\right)}
 \def\[{\left[}
\def\]{\right]}

\newcommand*{\beq}{\begin{equation}}
\newcommand*{\eeq}{\end{equation}}
\newcommand*{\bea}{\begin{eqnarray}}
\newcommand*{\eea}{\end{eqnarray}}

\newcommand \hrho{\hat{\rho}_1}

 \newcommand{\virgolette}[1]{``#1''}
 
\definecolor{ao(english)}{rgb}{0.0, 0.5, 0.0}
\definecolor{forestgreen(web)}{rgb}{0.13, 0.55, 0.13}

\DeclareMathOperator{\e}{e}

\def\simge{\mathrel{%
       \rlap{\raise 0.511ex \hbox{$>$}}{\lower 0.511ex \hbox{$\sim$}}}}
\def\simle{\mathrel{
       \rlap{\raise 0.511ex \hbox{$<$}}{\lower 0.511ex \hbox{$\sim$}}}}

\begin{document}

\title{On the Properties of a Bundle of Flexible Actin Filaments in an Optical Trap. }

\author{Alessia Perilli}
\email{alessia.perilli@roma1.infn.it}
\affiliation{Department of Physics, Sapienza University of Rome, P.le Aldo Moro 2, I-00185 Rome, Italy}
\author{Carlo Pierleoni}
\email{carlo.pierleoni@aquila.infn.it}
\affiliation{Department of Physical and Chemical Sciences, University of L'Aquila, Via Vetoio 10, 67100  L'Aquila, Italy}
\author{Giovanni Ciccotti}
\email{giovanni.ciccotti@roma1.infn.it}
\affiliation{Department of Physics, Sapienza University of Rome, P.le Aldo Moro 2, I-00185 Rome, Italy\\
School of Physics, University College Dublin (UCD), Belfield, Dublin 4, Ireland}
\author{Jean-Paul Ryckaert}
\email{jryckaer@ulb.ac.be}
\affiliation{Department of Physics, Universit\'e Libre de Brussels (ULB), Campus Plaine, CP 223, 
B-1050 Brussels, Belgium\\
Department of Physical and Chemical Sciences, University of L'Aquila, Via Vetoio 10, 67100  L'Aquila, Italy}

\date{\today}

\begin{abstract}
We establish the Statistical Mechanics framework for a bundle of $N_f$ living and uncrosslinked actin filaments in a supercritical solution of free monomers pressing against a mobile wall. The filaments are anchored normally to a fixed planar surface at one of their ends and, because of their limited flexibility, they grow almost parallel to each other. Their growing ends hit a moving obstacle, depicted as a second planar wall, parallel to the previous one and subjected to a harmonic compressive force. The force constant is denoted as trap strength while the distance between the two walls as trap length to make contact with the experimental optical trap apparatus.
For an ideal solution of reactive filaments and free monomers at fixed free monomers chemical potential $\mu_1$, we obtain the general expression for the grand potential from which we derive averages and distributions of relevant physical quantities, namely the obstacle position, the bundle polymerization force and the number of filaments in direct contact with the wall. 
The grafted living filaments are modeled as discrete Wormlike chains (d-WLC), with F-actin persistence length $\ell_p$, subject to discrete contour length variations $\pm d$ (the monomer size) to model single monomer (de)polymerization steps. 
Rigid filaments ($\ell_p=\infty$), either isolated or in bundles, all provide average values of the stalling force in agreement with Hill's predictions $F_s^H=N_f k_BT\ln(\rho_1/\rho_{1c})/d$, independent of the average trap length. Here $\rho_1$ is the density of free monomers in the solution and $\rho_{1c}$ its critical value at which the filament doesn't grow nor shrink in the absence of external forces. Flexible filaments ($\ell_p<\infty$) instead, for values of the trap strength suitable to prevent their lateral escape, provide an average bundle force and an average trap length slightly larger than the corresponding rigid cases (few percents). Still the stalling force remains nearly independent on the average trap length, but results from the product of two strongly $L$--dependent contributions: the fraction of touching filaments $\propto\left(\<L\>^{O.T.}\right)^2$ and the single filament buckling force $\propto\left(\<L\>^{O.T.}\right)^{-2}$.
\end{abstract}

% 61.25.he Polymer solutions
% 65.20.De 
%     General theory of thermodynamic properties of liquids, including computer
%     simulation
% 82.35.Lr Physical properties of polymers
%87.14.em	Fibrils (amyloids, collagen, etc.)
% 87.15.La	Mechanical properties of biomoleculesL

\maketitle

\section{Introduction}
Eukaryotic cells in biological environments are able to store chemical energy in ATP complexes and, by hydrolysis,
convert it into mechanical work used to perform several functions, e.g. movement and division. 
In particular, assembly and disassembly of actin microfilaments and microtubules are one of the main fundamental
 processes in the cells which produce mechanical forces against obstacles, such as membranes or bacteria: 
 filaments with one end anchored to the cytoskeletal network (pointed end)
 and with the growing end (barbed end) pointing toward 
 the obstacle, polymerize and depolymerize while staying in contact with the obstacle and pushing it away. Actin 
 filaments in cells are usually organized into fairly rigid bundles with the help of fascin, an actin cross-linking 
 protein, while their growth is controlled by capping proteins, which prevent them from becoming too long and 
 flexible \cite{Howard}; due to these features and to the intrinsic large stiffness of these filaments, most of the existing 
 models discard their flexibility and treat them as infinitely stiff. In this paper, following the lines set by \cite{Frey, Ryckaert, Paper1}, we investigate the role of flexibility in the process of reversible work production by F-actin filaments. Over the last decades, the underlying mechanism which enables cells to produce forces has been extensively studied, both theoretically and experimentally.
 
Originally, in the early 80's, a purely thermodynamic approach was followed by T.L. Hill \cite{Hill,Hill2}. The system under study is an almost incompressible (1D) polymer, actually a linear structure of length $L =L(N,F,T)$ consisting of $N$ self-assembled monomers of size $L/N\approx l_0$ at temperature $T$ which is confined by a compressive force $F$. By considering a supercritical chemical equilibrium between the confined polymer's monomers and a free monomer solution with monomer chemical potential $\mu_1$ at density $\rho_1$, Hill showed that $F=k_BT\ln\hrho/l_0$, where 
\begin{equation}
\hrho=\frac{\rho_1}{\rho_{1c}}=\exp\left(\beta\left(\mu_1-\mu_{1c}\right)\right)
\label{rho1}
\end{equation}
and the critical state, with free monomer solution density $\rho_{1c}$ and chemical potential $\mu_{1c}$, corresponds to the thermodynamic state in which the same polymer has a propensity neither to grow nor to shrink in the absence of any external force. If the formula is adapted to a bundle of $N_f$ rigid and parallel actin filaments, one gets a stalling force given by
\begin{equation}
\label{F_hill}
F^H_s=N_f\frac{k_BT}{d}\ln\hat\rho_1   
\end{equation}
For F-actin $d=2.7~nm$ is the contour elongation due to the addition of a subunit.\footnote{In the case of actin filaments, which consist of two interwoven protofilaments shifted with respect to each other by a distance equal to half the size of a G-actin (globular actin) monomer, $d$ corresponds to half its globular diameter. The linear self-assembled microfilament is called F-actin. We will make use of this terminology along the paper.}

Later Brownian Ratchet models (BRM) \cite{Peskin, Mogilner, MO.99, Mogilner2, Mogilner3, sander.00, Tsekouras2011, Demoulin} have been formulated in order to provide a more mechanistic interpretation of the action of a bundle of rigid filaments against a loaded obstacle. The ratcheting mechanism is played by the intercalation of a monomer between the filament tip and the pushing barrier whenever thermal fluctuations of the obstacle open a gap between them wide enough to allow for a polymerization event to occur. Attempts to include filament tip flexibility by adding a supplementary ratcheting mechanism have been published some years ago\cite{MO.99,Mogilner2}. Under the hypothesis that thermal fluctuations of the obstacle are fast compared to the frequency at which monomers attach/detach, the Brownian Ratchet models for rigid filaments provide a value of the stalling force in agreement with Hill's prediction, irrespectively of the disposition of the filaments seeds \cite{Peskin, sander.00, Tsekouras2011, Demoulin}. 

From an experimental perspective, the determination of the stalling force could be realized, in principle, by interpolating/extrapolating data of the bundle growing or shrinking velocity versus load \cite{Dogterom,Demoulin,Cojoc2007,Greene2009,Banquy2013,Farrell2013} and determining the zero velocity conditions. This route is in practice very difficult to follow given the noise level and interferences with hydrolysis of ATP-actin. Hence, 
Footer et al. \cite{Footer} used an optical trap set-up to measure the forces generated by the elongation 
of a few parallel-growing actin filaments in contact with a rigid microfabricated barrier, equilibrium being established between the bundle polymerization force and the trap restoring force directly proportional to the trap length. We observe that this set up represents in principle a true stable equilibrium state, as long as temperature and free monomer chemical potential are kept fixed and the implied chemical reactions are reversible with no filament escaping laterally along the obstacle wall \cite{Footer, Marenduzzo, Paper1}. Footer et al. monitored the growth of approximately eight actin filaments and found a stationary force significantly smaller than the value predicted by Hill's theory. The force measured in this experiment was of the order of the $F^{H}_{s}$ from Eq.(\ref{F_hill}) expected for a single filament: the interpretation of this unique (as far as today) and important experiment probing stalling conditions is still missing even if some possibles causes have been evoked \cite{Footer}. In ref. \cite{Carlsson2008} Carlsson investigated the effects of hydrolysis and irreversible conversion of ATP-complexed in ADP-complexed actin monomers on stalling conditions, within the framework of the BRM with a $L$ dependent load. Still considering fully rigid filaments, he found that the hydrolysis could account for the experimental observation in Footer's experiments. However due to the lack of experimental informations about the ADP off rate, the theoretical predictions remain inconclusive. Flexibility effects could also give rise to a decrease of the bundle force at stalling as observed in experiments because the bundle can ``buckle''. This was invoked by the authors to justify the results \cite{Footer}, although the arguments remained rather qualitative. Along these lines, in a simulation approach of filament growth against a constant load \cite{Marenduzzo}, flexibility was found to prevent the establishing of a true stationary non-equilibrium state since beyond some length, semi-flexible filaments can loose contact with the wall and grow parallel to it, hence reducing the force they are able to provide. This phenomenon has been called ``pushing catastrophe" in the context of constant-load experiments \cite{Marenduzzo} and ``escaping filaments" in a study restricted to equilibrium conditions \cite{Paper1}.

Recently, the force exerted by Brownian fluctuations of a grafted semi-flexible polymer, modelled as a Wormlike chain (WLC)
with fixed contour length, upon a rigid wall has been calculated both analytically and by Monte Carlo method, finding a force, entropic in origin, which exhibits a universal behavior in the stiff limit \cite{Frey}. 
The discrete version of this model (d-WLC) has then been extended to the case of a bundle of independent \virgolette{living} filaments growing in contact with a rigid fixed wall in a reactive canonical ensemble \cite{Ryckaert}; within this statistical mechanical description several general features have been derived, namely the equilibrium filament size distribution and the associated average equilibrium force exerted on the opposite wall. Along these lines, a recent study \cite{Paper1} has extended this analysis to the reactive grand canonical ensemble, specified by temperature $T$, volume $V$ and free monomers chemical potential $\mu_1$, for a single grafted living semi-flexible filament modeling F-actin, hitting a fixed wall. 

The natural extension towards the properties of a bundle of parallel semi-flexible actin filaments pressing against a mobile loaded wall is the subject of the present work. As already mentioned we limit here to equilibrium statistical mechanics and we consider an external load increasing with the position of the obstacle in order to focus on a true equilibrium state. We will consider the case of a load increasing linearly with $L$, to mimic the experimental relevant case of a bundle in a harmonic optical trap \cite{Footer}. Furthermore, we disregard the ATP-ADP conversion through hydrolysis which introduces an inherent irreversible process hence a non-equilibrium situation which remains to be studied. Within our approach, we establish the physical conditions to avoid the escaping filaments regime for flexible filaments, a task that \textit{in vivo} is performed by specific proteins (capping, fascin). We characterize the effects of flexibility at equilibrium by comparing relevant properties, such as the average trap width, bundle force and number of active filaments, for a F-actin bundle using either flexible or rigid filaments. Moreover, the statistical mechanics foundations of Hill's expression, Eq.(\ref{F_hill}), of the stalling force are analyzed in depth.

The paper is organized as follows. 
In section \ref{sec2} we define the physical system 
and, in particular, the homogeneous and in--registry bundles of living filaments based on the disposition of their seeds; we set-up the statistical mechanics framework for a bundle of filaments in the fixed--wall reactive grand canonical ensemble, and derive the expression of the average relevant properties of the system. 
In section \ref{sec3} the moving-wall ``ensemble" is introduced for a restoring hookean force acting on the wall to make contact with the experiment realized by Footer et al. \cite{Footer}. We close the section defining the meaning of force measurement in terms of ensemble averaging.
In section \ref{sec4} we briefly recall our model of F-actin filaments \cite{Paper1, Frey} and extend the non--escaping filament regime criteria of ref. \cite{Paper1} to the present case of the optical trap.
Section \ref{sec5} presents our results. We first characterize the flexibility effects by comparing rigid and flexible models for single filaments and then for a bundle of filaments. We also characterize the behavior of flexible filaments bundles in a wide range of parameters and physical conditions, going from the quasi--rigid filament behavior for short filaments to near the threshold of the escaping regime, revealing an intermediate regime characterized by a growing cooperativeness between filaments to produce the equilibrium force. 
In section \ref{sec6} we discuss and suggest an explanation for the experimental results \cite{Footer} and draw few conclusive remarks.

\section{Bundle of living, supercritical grafted filaments in a box} \label{sec2}

We consider a bundle of $N_f$ independent (mutually non-interacting) stiff filaments enclosed in a box of constant transverse area $A$ and constant height $L$ with two parallel and opposite walls located at $x=0$ and $x=L$; the filaments, according to the discrete WLC model (material points and bonds) with bond length $d$ and persistence length $l_p$, are anchored normally to the first wall and can grow towards the second wall. We consider the obstacle at $L$ as a hard wall, i.e. no filament articulation point (in particular the filament tip) can overlap the wall region beyond $L$. The filaments are immersed in an ideal solution of free monomers (material points not interacting with each other) at chemical potential $\mu_1$. Single monomer polymerization and depolymerization events give to the filaments their living character with probabilities satisfying chemical equilibrium. 
We showed \cite{Paper1} that, as a result of the chemical equilibrium,
the free energy total differential of our confined system can be expressed as 
\begin{equation}
\label{dOmega}
d\Omega^R=-SdT-p_NA dL-p_TLdA-N_td\mu_1+\left(\mu_2-2\mu_1\right)dN_f
\end{equation}
where $S$ is the system entropy, $p_N$ and $p_T$ the total normal and tangential pressures exerted by the wall on the system, $\mu_2$ is the chemical potential of grafted dimers, $N_t=N_1+\sum_{n=1}^{N_f}j_n$ is the total number of particles (free plus bonded monomers) and $\mu_1$ is the chemical potential of the free monomers, which results conjugated to $N_t$ as a consequence of chemical equilibrium. The last two terms on the right hand side arise from iteratively applying to the free energy differential of a mixture of all chemical species the equilibrium condition, $\mu_{i+1}=\mu_i+\mu_1$, where $\mu_i$ is the chemical potential of the grafted filament of size $i$.
We are interested in supercritical conditions, to be defined more precisely below, when the filaments tend to grow in bulk (polymerization rate greater than depolymerizing rate) but reach an equilibrium as a result of the obstacle wall capacity to reduce the polymerization rate of hitting filaments.

The longitudinal disposition of the filament seeds (first two monomers) at the grafting surface represents a significant characteristic for a bundle. Its influence on the structural properties of the systems, often discussed within the context of multi-filament brownian ratchet models \cite{Peskin, Mogilner, Mogilner2}, will be discussed in the next sections. In absence of an experimental information we limit our analysis to the two usually adopted models: we call a bundle \textit{homogeneous} when the seeds are regularly distributed over a distance $d$ centered at $x=0$ (the position of the grafting wall) while we call a bundle \textit{in--registry} when all seeds are aligned at $x=0$. 
Labelling $h_n$ the longitudinal position of the seed of the $n$--th filament, we set
\begin{equation}
\label{h_n}
h_n = \begin{cases} \left(\frac{n-0.5}{N_f} - 0.5\right)d &\mbox{homogeneous bundle} \\ 
0 & \mbox{in-registry bundle}  \end{cases} \hskip 2cm n\in [1,N_f]
\end{equation}
The distance between the first monomer of filament $n$ and the wall at $x=L$ is given by $L_n=L - h_n$. Following notations of ref. \cite{Paper1}, the contour length of a filament of $j$ monomers is $L_{c,j}=(j-1)d$ where $d$ is the bond length. 
The minimum number of monomers in a filament is taken to be two, at least two monomers are needed to specify the growth 
direction kept perpendicular to the transverse surface $A$. The maximum number of monomers in a filament with its first monomer at $x=h_n$, before it feels the presence of the obstacle at $L$ is 
\begin{equation}
\label{z_n}
z_n=int\left(\frac {L_n}d\right) + 1 
\end{equation}
The second critical filament size index introduced in \cite{Paper1}%Another important index is
\begin{equation}
\label{z*}
z^\ast_n = int\left(\frac{\pi L_n}{2d}\right)
\end{equation}
corresponds to a contour length equal to a quarter of a circle of radius $L_n$; a filament with a number of monomers larger than 
$z^\ast$ is considered an \textit{escaping filament} since in supercritical conditions, for planar conformations, it may grow unhindered in the direction parallel to the obstacle\cite{Paper1}.

Let $q_{j_n}(L_n)$, and $q_{j_n}^0$ be the partition functions of a single grafted filament $n$ (with seed at $x=h_n$) having size $j_n$, respectively in presence and in absence of the wall. We define the wall factors $\alpha(j_n,L_n)$ of each specific filament as:
\begin{equation}
\label{z}
\alpha(j_n,L_n)=\frac{q_{j_n}(L_n)}{q_{j_n}^0}\begin{cases} = 1 & 2\leqslant\ j_n\leqslant z_n\\ 
< 1 & z_n<  j_n<z_n^\ast \end{cases} 
\end{equation}

The size-independent chemical equilibrium constant for the (de)polymerization reaction in the bulk system is, considering two grafted filaments (i.e. their $q_i^0$ have not to be divided by the volume, see Eq.(10.6) of ref.\cite{Hill_book}) of arbitrary sizes $i-1$ and $i$,
\begin{equation}
\label{K0}
K_0=\frac{q_{i}^0}{q_{i-1}^0q_1/V}=\Lambda^3 \frac{q_{i}^0}{q_{i-1}^0} \qquad i=2,z^\ast
\end{equation}
where $q_1=V/\Lambda^3$ is the free monomer partition function and $\Lambda(T)=\sqrt{\beta h^2/2\pi m}$ is the free monomer thermal de Broglie wavelength. Combining Eqs.(\ref{z},\ref{K0}), the filament partition can be expressed as
\begin{equation}
\label{ki}
q_{j_n}=q_2^0~\alpha(j_n,L_n) \left(\frac{K_0}{\Lambda^3}\right)^{(j_n-2)}
\end{equation}

The explicit expression of the grand--canonical partition function for a single non escaping living filament in the reactive-grand-canonical ensemble has been derived in ref. \cite{Paper1} (see Eqs.(6-31) of that paper). The extension to a bundle of independent living filaments is straightforward: we need to sum over all possible $N_t$ the canonical partition function, $Q^R(A,L,T,N_t,N_f)$, involving $N_t$ monomers and $N_f$ filaments, properly weighted by the corresponding absolute activities, 
\bea
\Xi^R(A,L,T,\mu_1,N_f)&=&\sum_{N_t=2N_f}^{\infty}e^{\beta\mu_1N_t}Q^R(A,L,T,N_t,N_f)\nonumber\\
&=&\sum_{N_t=2N_f}^{\infty}\ e^{\beta\mu_1 N_t}\sum_{\mathclap{\substack{j_1=2\\
\qquad\qquad N_t=N_1+\sum_{n=1}^{N_f}j_n}}}^{z^*_1}\dots\sum_{j_{N_f}=2}^{z^*_{N_f}}\ \frac{q_1^{N_1}}{N_1!}\ q_{j_1}(L_1)\dots q_{j_{N_f}}(L_{N_f})\\
&=&\sum_{N_1=0}^{\infty}\sum_{j_1=2}^{z^*_1}\dots\sum_{j_{N_f}=2}^{z^*_{N_f}}\  e^{\beta\mu_1 N_t} \  \frac{q_1^{N_1}}{N_1!}\ q_{j_1}(L_1)\dots q_{j_{N_f}}(L_{N_f}).
\label{eq:QR}
\eea 

The grand-canonical partition function Eq.(\ref{eq:QR}) can be further expressed as the product of single filaments and free monomers partition function. Indeed using Eqs.(\ref{z},\ref{eq:QR}) one gets:
\bea
\label{eq:QR1}
\Xi^R(A,L,T,\mu_1,N_f)&=&\sum_{N_1=0}^{\infty} \frac{q_1^{N_1}}{N_1!} \ e^{\beta\mu_1 N_1}  \prod_{n=1}^{N_f} \left[ \sum_{j_n=2}^{z^*_n}q_{j_n}(L_n) e^{\beta\mu_1 j_n}\right]\\
&=&\Xi^{free}(A,L,T,\mu_1) \left(\frac{q_2^0 \Lambda^6}{K_0^2}\right)^{N_f} \prod_{n=1}^{N_f} \left[\sum_{j_n=2}^{z^*_n} \alpha(j_n,L_n) \hat{\rho}_1^{j_n} \right]
\eea
where $\rho_1=e^{\beta\mu_1}/\Lambda^3$ since the monomers are a perfect gas, $\rho_{1c}=K_0^{-1}$ and $\hrho=\rho_1/\rho_{1c}$ and $\Xi^{free}(A,L,T,\mu_1)$ is the free monomer ideal gas partition function in the accessible volume at same temperature and chemical potential. The free energy $\beta\Omega^R=-\ln{\Xi^R}$ takes the form
\bea
\beta\Omega^R(A,L,T,\mu_1,N_f)&=&\beta\Omega^{free}(A,L,T,\mu_1) +\beta\Omega^{bun}(L,T,N_f,\mu_1)\label{eq:OmegaR}\\
\beta\Omega^{free}(A,L,T,\mu_1)&=&-\frac{pV}{k_BT}=-\frac{AL\hrho}{K_0}\label{eq:Omegafree}\\
\beta\Omega^{bun}(L,T,\mu_1,N_f)&=&-N_f\ln\left(\frac{q_2^0\Lambda^6}{K_0^2}\right)-\sum_{n=1}^{N_f} \ln \mathcal{D}(L_n,\hrho) \label{eq:Omegabun}
\eea
which is the natural generalization of the single filament case in ref. \cite{Paper1}. In the r.h.s. of Eqs (\ref{eq:Omegafree},\ref{eq:Omegabun}), the $\mu_1$ dependence follows from the link between $\hat\rho_1$ and the free monomer chemical potential given in Eq.(\ref{rho1}). 
The use of $\hat\rho_1$ instead of $\mu_1$ is very common in the biophysics literature as it has a direct interpretation as the ratio of polymerization and depolymerization rates in the bulk\cite{Paper1} and as it allows more compact expressions. 
This applies in the last term in the bundle free energy Eq.(\ref{eq:Omegabun}) where we have introduced the partition function $\mathcal{D}(L_n,\hrho)$ of the single living filament of index $n$
\begin{equation}
\label{D}
\mathcal{D}(L_n,\hrho)=\sum_{j_n=2}^{z^\ast_n}\alpha(j_n,L_n)\hat\rho_1^{j_n}.
\end{equation}
This partition function is directly linked to the probability
\begin{equation}
\label{P(j)}
\mathcal{P}(j_n|L_n,\hat\rho_1)=\frac{\alpha(j_n,L_n)\hat\rho_1^{j_n}}{\mathcal{D}(L_n,\hat\rho_1)} \ \ \ \ \ \ \ j_n=2,z^\ast_n
\end{equation}
for filament of index $n$ to have a size $j_n\in[2,\ z^\ast_n]$ given the seed-wall distance $L_n$ and the reduced density $\hrho$ \cite{Ryckaert,Paper1}.

The knowledge of the set of $\mathcal{D}(L_n,\hat\rho_1)$ for the filaments in the bundle allows to compute all equilibrium properties of the bundle. Moreover our notations allow to treat both flexible and rigid filaments: for the rigid case 
\begin{equation}
\alpha(j_n,L_n)=\begin{cases}1& j_n\leqslant z_n\\0& j_n>z_n
\end{cases}\qquad\qquad \mathcal{D}(L_n,\hrho)=\sum_{j_n=2}^{z_n}\hrho^j=\frac{\hrho^2}{1-\hrho}\left(1-\hrho^{|\frac Ld|}\right)
\label{Drig}
\end{equation}
where $|\cdots|$ means the integer part of the argument.

According to Eq.(\ref{dOmega}), the partial derivative of the grand potential with respect to $L$ gives the 
total normal pressure exerted on the wall. Using Eqs.(\ref{eq:Omegafree},\ref{eq:Omegabun}) we thus get
\beq
p_N = -\frac1A\frac{\partial\Omega^R}{\partial L} =k_BT\frac{\hrho}{K_0}+ \frac{1}{A}F_{bun}\left(L,\hat\rho_1\right) 
\label{Fbun1}
\eeq
where 
\bea
F_{bun}\left(L,\hat\rho_1\right) & = &  -\frac{\partial\Omega^{bun}}{\partial L} = k_BT\sum_{n=1}^{N_f}\frac{\partial\ln\mathcal{D}(L_n,\hat\rho_1)}{\partial L_n}\nonumber\\
\label{Fbun}
&=&k_BT\sum_{n=1}^{N_f}\sum_{j_n=z_n+1}^{z^{\ast}_n}\frac{\alpha(j_n,L_n)\hat\rho_1^j}{\mathcal{D}(L_n,\hat\rho_1)}\frac{\partial \ln{\alpha(j_n,L_n)}}{\partial L_n}\\
&=&k_BT\sum_{n=1}^{N_f}\sum_{j_n=z_n+1}^{z^{\ast}_n} \mathcal{P}(j_n|L_n,\hat\rho_1) \frac{\partial \ln{\alpha(j_n,L_n)}}{\partial L_n}.
\end{eqnarray}
Introducing $f_{\bot}(L_n,\hrho)$, the equilibrium force exerted by the $n^{th}$ living filament of the bundle on the wall, we can write:
\bea
F_{bun}\left(L,\hat\rho_1\right) &=& \sum_{n=1}^{N_f}f_{\bot}(L_n,\hrho)  \\
\label{f_N}
f_{\bot}(L_n,\hat\rho_1)&=&\sum_{j_n=z_n+1}^{z^\ast_n}\mathcal{P}(j_n|L_n,\hrho)\overline{f}_{j_n}(L_n)
\eea
where $\overline{f}_j(L_n)=-\partial W_j(L_n) / \partial L$, with $W_j(L_n)=-k_BT[\ln\alpha(j,L_n)]$ the corresponding potential of mean force related to the presence of the wall, is the mean force exerted by a filament of fixed size $j$ on the wall distant $L_n$ from its seed. Note that for rigid filaments the concept of force becomes ill-defined since the potential of mean force of a filament of contour length $L_{c,j}$ is either zero for $L_n\geq L_{c,j}$ or infinite for $L_n<L_{c,j}$. Correspondingly $\mathcal{P}(j_n|L_n,\hrho)$ goes to zero for $L_n<L_{c,j}$.

The partial derivative of $\Omega^R$ with respect to $\mu_1$, gives the average number of monomers in the system 
\begin{eqnarray}
N_t(L,\hrho) & = & AL\frac{\hat\rho_1}{K_0(T)} + \sum_{n=1}^{N_f}\frac{\partial\ln\mathcal{D}(L_n,\hat\rho_1)}{\partial\ln\hat\rho_1} \nonumber\\
&=&  AL\rho_1 -\frac{\partial\beta\Omega^{bun}}{\partial\ln\hat\rho_1} = AL\rho_1+N_f L_{bun}(L,\hrho)
\eea
where we used $\hrho=K_0e^{\beta\mu_1}\Lambda^{-3}$ and 
\bea
L_{bun}(L,\hrho) & = & -\frac1{N_f}\frac{\partial\beta\Omega^{bun}}{\partial\ln\hat\rho_1}\nonumber\\
\label{Lbunav}
&=&\frac1{N_f}\sum_{n=1}^{N_f}\sum_{j_n=2}^{z_n^\ast}j_n\mathcal{P}(j_n|L_n,\hrho)
\end{eqnarray}
is the average bundle size.

To define the number of filaments hitting a wall, it is appropriate to define a new filament relative--size probability, $Q(m_n|L,\hrho)$, relative to the distance from the wall position, where the index $m_n\equiv j_n-z_n$ runs in the interval $m_n\in[2-z_n, z^\ast_n-z_n]$:
\beq
Q(m_n|L,\hrho)=Q(j_n-z_n|L_n,\hrho)=P(j_n|L_n,\hrho)
\label{eq:Q_mn}
\eeq

In terms of the absolute-- and relative--size distributions, the expected total number of filaments $N_0$ touching the fixed obstacle at given position $L$, is the sum of the probability for each filament of the bundle to have a size larger than $z_n$, hence
\bea
\label{N_0}
N_0(L,\hrho)&=&\sum_{j_1,\dots, j_{N_f}=2}^{z^\ast_n}\left[\prod_{n=1}^{N_f}\mathcal{P}(j_n|L_n,\hrho)\right]\sum_{n=1}^{N_f}\Theta(j_n-z_n-1)\nonumber\\
\label{eq:n0}
&=&\sum_{n=1}^{N_f}\sum_{j_n=z_n+1}^{z_n^\ast}\mathcal{P}(j_n|L_n,\hrho)=\sum_{n=1}^{N_f}\sum_{m_n=1}^{k^\ast} Q(m_n|L_n,\hrho)
\eea
where $\Theta(j_n-z_n-1)$ is an indicator which is unity if the argument of $\Theta$ is non negative and zero if it's negative.
In the case of rigid filaments, $N_0$ vanishes.  

\section{Optical Trap Ensemble}
\label{sec3}
\begin{figure}
\begin{center}
\includegraphics[totalheight=5cm]{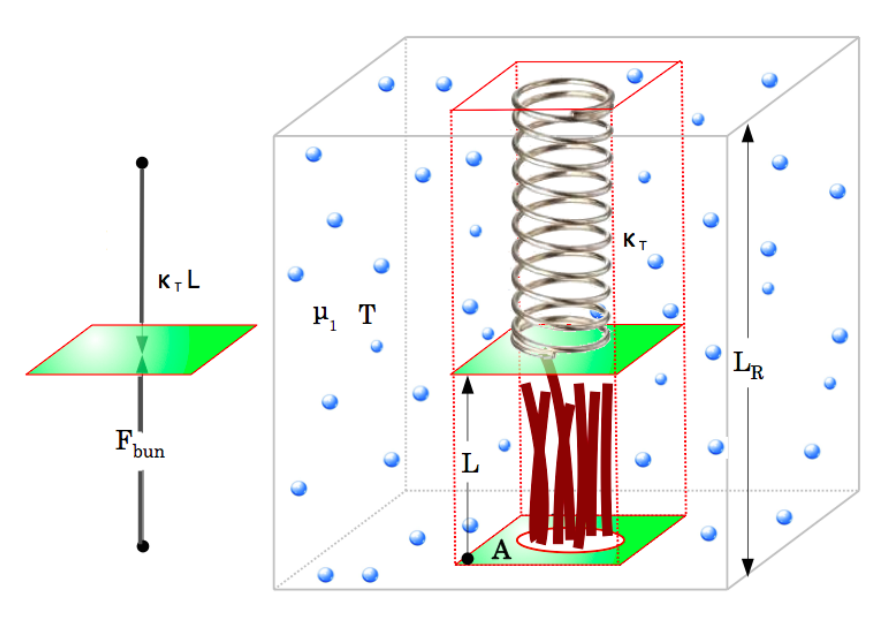}%width=0.5\textwidth
\caption{In this representation of the optical trap set-up, the colloid used in the experiment, monitored during the 
filaments growth, is pictured as a horizontal moving wall connected to a spring of elastic constant $\kappa_T $. The filaments are 
grafted to a horizontal fixed wall and grow against the moving one, exerting an upwards force. The compressed spring, 
with physical rest length equal to $L_R$, exerts a downwards force $\kappa_T L$, where $L$ is the distance between the moving and the 
grafting planes. The moving wall divides the total volume 
$V_{tot}=L_RA$ into two sub-volumes, which are in global equilibrium at a 
fixed temperature $T$ and are surrounded by a large reservoir containing a solution of G-actin (represented by little blue 
spheres) at fixed chemical potential $\mu_1$ and fixed density $\hat{\rho}_1$. }
\label{OT_fig}
\end{center}
\end{figure}

Footer et al. \cite{Footer} measured the force exerted by a bundle of approximately eight F-actin filaments by opposing to the growing filaments a colloidal particle subjected to a restoring force. The force, linear in the colloid displacement from its position in absence of the bundle, was generated by trapping the colloid within an optical trap apparatus and was indirectly measured by monitoring the displacement of the colloid during the growth of the bundle. For large enough time the colloid position reached a stationary state because of the harmonic restoring force. Figure \ref{OT_fig} schematically shows an equivalent set-up. 
In a large volume filled with free monomers at fixed chemical potential $\mu_1$ and hence at fixed grand--canonical average density $\hrho$, and in contact with a heat bath at temperature $T$, consider a central volume defined by a transverse area $A$ and a fixed length $L_R$.
The central volume of size $V=A L_R$ is bounded on one side by a fixed wall of area $A$ into which the filaments are grafted. Additionally, this volume is partitioned into two chambers of common transverse area but of variable heights, by a mobile hard wall parallel to the grafting wall which can only move vertically. Chamber $I$ with length $L$ encloses some free monomers and the bundle with filaments pressing on the moving wall which is further subjected to a restoring force $\kappa_T L$ (represented by the spring in chamber $II$) modeling the trapping mechanism affecting the colloidal particle. Chamber $II$ of complementary length $L_R-L$ contains only free monomers which exert some pressure on the separating wall. 
The moving wall here represents the colloidal particle used in the experiment \cite{Footer}. The total free energy of this system is given by:
\bea
\Omega^{O.T.}(A,&L&,L_R,T,N_f,\mu_1,\kappa_T)=\Omega_1(A,L,T,N_f,\mu_1)\nonumber\\
\label{Omega_ot}
&+&\Omega_2(A,L_R-L,T,\mu_1)+\frac12\kappa_T L^2
\label{O_OT}
\eea
where $\Omega_1(A,L,T,N_f,\mu_1)$ is the grand potential of the first sub-volume containing the bundle of $N_f$ filaments, 
$\Omega_2(A,L_R-L,T,\mu_1)$ is that of the second sub-volume containing only the solution of free monomers and 
$\frac{1}{2} \kappa_T L^2$ is the contribution due to the compressional energy stored in the spring.

Eq.(\ref{Omega_ot}) is valid at thermodynamic equilibrium and it will be applicable to experimental situations when this condition is satisfied.
The grand potential for the ideal free monomer solution is 
\beq
\label{Omega2}
\beta\Omega_2(A,L_R-L,T,\mu_1)=-\frac{A\(L_R-L\)}{K_0}\hat\rho_1
\eeq
while the grand-potential of the first chamber is given by Eq.(\ref{eq:OmegaR}). 
Summing these terms to the elastic free energy the expression for the optical trap grand potential is
\bea
\Omega^{O.T.}(L,T,N_f,\mu_1,\kappa_T)&=&\frac12\kappa_T L^2+\Omega^{bun}(L,T,N_f,\mu_1) -\frac{AL_R}{\beta K_0}\hrho\nonumber\\
&=& \frac12\kappa_T L^2 - k_BT\sum_{n=1}^{N_f}\ln\mathcal{D}(L_n,\hat\rho_1)-\frac{AL_R}{\beta K_0}\hrho-N_fk_BT\ln\left(\frac{q_2^0\Lambda^6}{K_0^2}\right)
\label{Omega_fin}
\eea

It is now convenient to define an equilibrium distribution for the variable $L$ through
\bea
&&P^{O.T.}(L|T,N_f,\mu_1,\kappa_T )\equiv P^{O.T.}(L)\nonumber\\
&=&\begin{dcases}
\frac{\exp\(-\beta\Omega^{O.T.}(L,T,N_f,\mu_1,\kappa_T )\)}{\int_{2d}^{L_R}dL^{\prime}\exp\(-\beta\Omega^{O.T.}(L^{\prime},T,N_f,\mu_1,\kappa_T)\)} = \frac{\exp\(-\frac{\beta \kappa_T L^2}{2}\) \left[\prod_{n=1}^{N_f} D(L_n,\hrho)\right]}{\int_{2d}^{L_R}dL^{\prime}\exp\(-\frac{\beta \kappa_T L^{\prime 2}}{2}\) \left[\prod_{n=1}^{N_f} D(L^{\prime}_n,\hrho)\right]}&\mbox{$2d<L<L_R$}\\0&\mbox{otherwise}
\end{dcases}
\label{eq:P(L)}
\eea

One can further define the joint distribution function $p(L,j_1,j_2,....j_{N_f}|\hrho)$ as 
\bea
p(L,j_1,j_2,....j_{N_f}|\hrho)= \frac{\exp\(-\frac{\beta \kappa_T L^2}{2}\) \left[\prod_{n=1}^{N_f} \alpha(j_n,L_n)\hrho^{j_n} \right]}{\int_0^{L_R}dL^{\prime}\exp\(-\frac{\beta \kappa_T L^{\prime 2}}{2}\) \left[\prod_{n=1}^{N_f} D(L^{\prime}_n,\hrho)\right]} = P^{O.T.}(L) \left[\prod_{n=1}^{N_f} \mathcal{P}(j_n | L_n,\hrho) \right]
\label{eq:P(Lj)}
\eea

The measurement of the force in the experiment \cite{Footer} has been obtained indirectly through that of the colloid position. In these conditions the measured force must be compared to the average over the $L$ distribution $P^{O.T.}(L)$, namely 
\beq
\<F_{bun}\>^{O.T.}=\int_0^{L_R}dL~P^{O.T.}(L)~F_{bun}(L,\hrho)
\label{Fexp}
\eeq
For the average colloid position one gets
\beq
 \<L\>^{O.T.}= \int_0^{L_R}dL~P^{O.T.}(L)~L.
\label{Lexp}
\eeq
From Eqs.(\ref{Fbun},\ref{Omega_fin}) we have $\kappa_TL=\frac{\partial\Omega^{O.T.}}{\partial L}+F_{bun}(L)$ and therefore 
\beq
\<L\>^{O.T.}=\frac1{\kappa_T}\int_0^{L_R}dL~P^{O.T.}(L)~\[\frac{\partial\Omega^{O.T.}}{\partial L}+k_BT\frac{\partial}{\partial L}\sum_{n=1}^{N_f}\ln\mathcal{D}(L_n,\hat\rho_1)\]
\eeq
where the first term vanishes noting that $P^{O.T.}(0)=P^{O.T.}(L_R)=0$. Therefore we have proved that $\<L\>^{O.T.}=\<F_{bun}\>^{O.T.}/\kappa_T$ showing that what is measured is equivalent to the optical trap average (i.e. an average over $L$) of the bundle force expression, as requested by mechanical equilibrium.

Of particular relevance is the optical trap, i.e. marginal, distribution of relative--size $Q$ introduced in Eq.(\ref{eq:Q_mn})\
\beq
Q^{O.T.}(m_n|\hrho)= \int_0^{L_R}dL~P^{O.T.}(L)~Q(m_n | L_n, \hrho)
\label{eq:Qm}
\eeq
from which we can compute the average fraction $\<x_0\>^{O.T.}=\frac{\<N_0\>^{O.T.}}{N_f}$ of touching filaments as 
\begin{equation}
\<x_0\>^{O.T.}=\frac1{N_f}\sum_{n=1}^{N_f}\sum_{m_n=1}^{z^\ast_n-z_n}Q^{O.T.}(m_n|\hrho)
\label{eq:x0}
\end{equation}

\section{F-Actin model}
\label{sec4}
\subsection{Dead filaments entropic force}
\label{sec4_a}
At the relevant length scales (a few microns at most) actin filaments are ``semiflexible polymers'' with large bending rigidity. 
The wall distance range of interest, $L\sim25d\div90d$, comparable to the filaments contour length $L_c$, is rather smaller than actin persistence length at room temperature $\ell_p=5370d$ ($d=2.7~nm$ is half of the size of the G-actin monomer) so that actin filaments behave as stiff chains.
As in ref. \cite{Paper1}, we adopt the living version of the discrete Wormlike Chain (d-WLC) model for F-actin. This model is particularly suited for our investigation since we can adopt the universal expression of Gholami et al.\cite{Frey} for the entropic force produced by ``dead filaments" as far as the filament contour length remains within the non-escaping regime (see ref. \cite{Paper1} and the following subsection). In the weak-bending ($L\lesssim L_c$), stiff-chain ($L_c\ll \ell_p$) regime the relevant adimensional variable is the \textit{reduced compression} \cite{Frey}
\beq
\label{eta}
\tilde\eta=\frac{\ell_p}{L_c}\(1-\frac{L}{L_c}\) \geqslant 0 \qquad\qquad (L\leqslant L_c)
\eeq
and the universal expression for the force-compression law of a continuous WLC of contour length $L_c$ is found to be
\beq
\label{f_j}
\overline{f}(L,L_c,\ell_p)=f_{b}(L_c,\ell_p)\tilde{f}_{\parallel}(\tilde\eta)
\eeq
where
\bea
\label{eq:fb_f||}
f_{b}(L_c,\ell_p)&=&\frac{\pi^2}4\frac{\ell_p}{L_c}\frac{k_BT}{L_{c}}\\
\label{f_tilde}
\tilde{f}_{\parallel}(\tilde\eta)&=&-\frac{4}{\pi^2}\frac{\partial\ln\alpha(\tilde\eta)}{\partial\tilde\eta}
\eea
and 
\beq
\label{alpha}
\alpha(j,L)\equiv\alpha(\tilde\eta)=\begin{cases}\sum_{k=1}^{\infty}\[\(-1\)^{k+1}\frac1{\lambda_k}\exp\(-\lambda_k^2\tilde\eta\)\]&\qquad \tilde\eta\geq0\\
1 &\qquad \tilde\eta<0
\end{cases}
\eeq
with $\lambda_k=\(2k-1\)\frac{\pi}{2}$.
$\tilde{f}_{\parallel}(\tilde\eta)$ starts from zero at $\tilde\eta=0$ ($L=L_c$) and rapidly increases with $\tilde\eta$ up to a unity plateau reached around $\tilde\eta=0.25$. Higher compressions do not increase the response force of this model. Note that the above behavior strictly concerns a continuos WLC. The extension of the WLC model to living filaments requires the use of a discrete WLC model whose contour length changes in a quantized fashion upon chemical events. 
As shown in ref. \cite{Paper1}, the use of the above theory for d-WLC model introduces negligible errors as far as the compression $\tilde\eta$ does not reach the breakdown of the weak-bending regime and the occurrence of the escaping regime for the living extension of the model. 

\subsection{Non-escaping filaments criteria}
\label{sec4_b}
Living filaments in supercritical conditions tend to grow indefinitely unless some external agent stops the preferential polymerization process. For continuously growing filaments we cannot define statistical equilibrium but at most a stationary non-equilibrium state. The growth of completely rigid filaments ($\ell_p\to \infty$) can always be arrested by a rigid obstacle provided the external force applied to the obstacle is strong enough to balance the bundle action. For semi-flexible filaments the situation is more complex because a filament that can bend, can also laterally escape and grow indefinitely because of the supercritical conditions. If this situation occurs we cannot use Equilibrium Statistical Mechanics to describe our system. In order to avoid the escaping state we have imposed that each filament in the bundle cannot have more than a maximum number of monomers $z^*_n= int\left(\pi L_n/2d\right)$ (see Eq.(\ref{z*})). Imposing a maximum number of monomers per filament, however, will bias the properties of the system unless the probability for $j_n=z^*_n$ to occur be negligibly small for all filaments in the bundle
\beq
\frac{\mathcal{P}(z^\ast_n|L_n,\hrho)}{\mathcal{P}(z_n|L_n,\hrho)}=\frac{\alpha(z^\ast_n,L_n)}{\alpha(z_n,L_n)}\hat\rho_1^{z^\ast_n-z_n}=\alpha(z^\ast_n,L_n)\hat\rho_1^{z^\ast_n-z_n}\ll1 \qquad\qquad \forall n \in[1,N_f]
\label{P_ratio}
\eeq 
since $\alpha(z_n,L_n)=1$.
Following ref. \cite{Paper1}, the non-escaping regime condition on the reduced density at fixed $L$ is found to be:
\beq
\hrho\exp\(-\frac{\ell_pd}{L_n^2}\)<1 \qquad\qquad \forall \ n \in[1,N_f]
\label{rho1lim}
\eeq
At fixed $\hrho$ this relation establishes a maximum amplitude of the box to avoid the $z^*$ bias. 
\begin{figure}
\begin{center}
\includegraphics[totalheight=8cm]{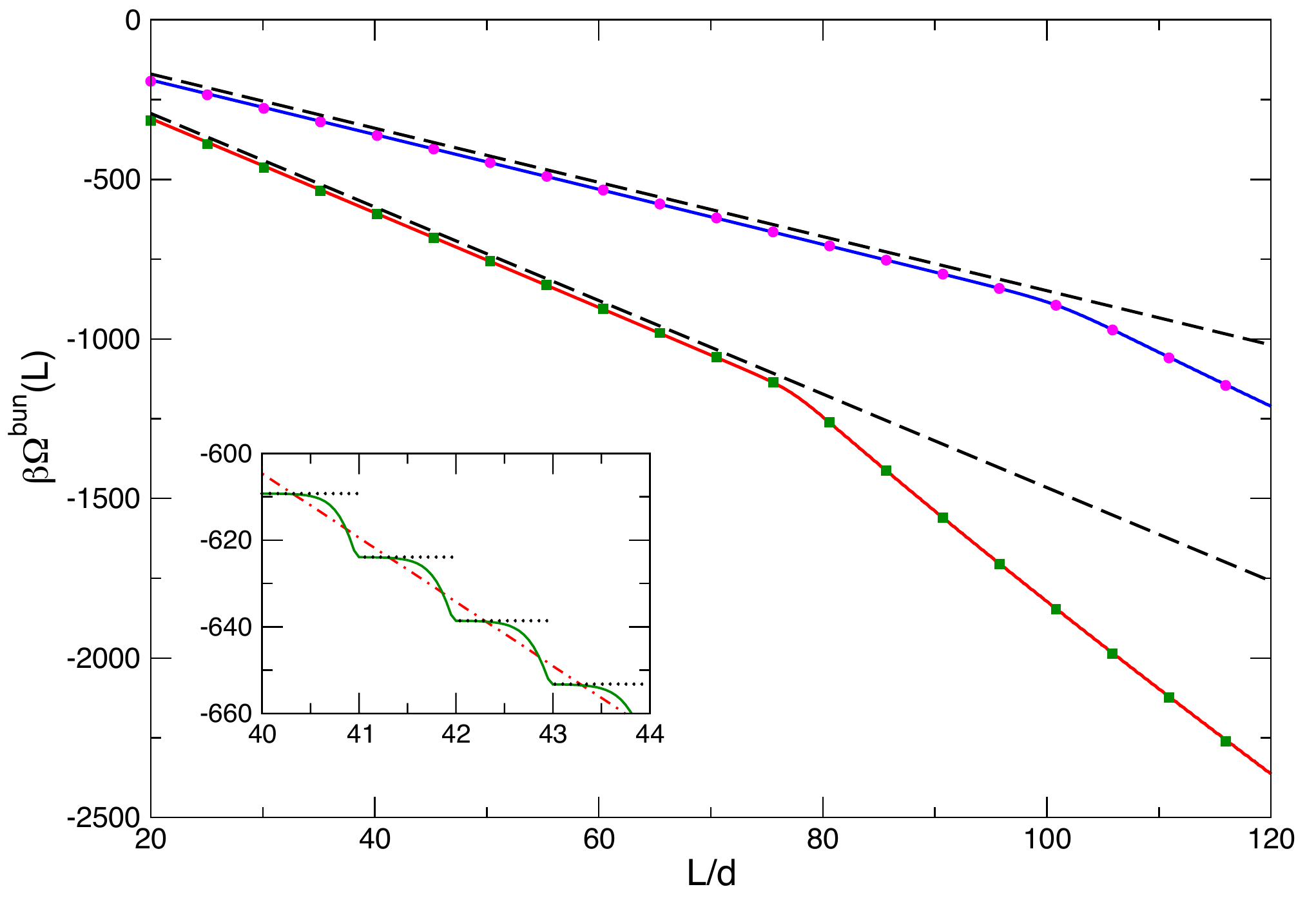}%width=0.6\textwidth
\caption{$L$ dependence of the bundle contribution to the grand-potential free energy 
$\beta \Omega^{bun}(L,t,\hrho,N_f)~+~N_f\ln\frac{q_20\Lambda^6}{K_0^2}~=~-~\sum_{n=1}^{N_f}\ln\mathcal{D}(\hat{\rho}_1,L_n)$ at room temperature for two free monomers reduced densities, $\hat{\rho}_1=1.7$ (blue continuous curve for homogeneous bundle, magenta closed circles for in registry bundle) and $\hat{\rho}_1=2.5$ (red continuous curve for homogeneous bundle, green closed squares for in registry bundle) and for bundles of $N_f=16$ filaments. Dashed lines represent, for both reduced free monomers densities, the linear function $- F_s^H L$ (from Eq.(\ref{F_hill})). The divergence from the dashed lines, for $L$ greater than $L_l=\sqrt{\ell_pd/\ln\hrho}$, corresponds to the breakdown of condition (\ref{rho1lim}) and then to the appearance of lateral escapes. Inset: magnification of the same data for $\hat{\rho}_1=2.5$ only for both the homogenous (red, dot-dashed line) and the in-registry (green continuous line) bundle. In limiting case of in-registry bundle of fully rigid filaments, this contribution to the free energy has a discontinuous pattern, as shown in the inset as black dots. See Eq.(\ref{Drig}).}
\label{fig:can_H}
\end{center}
\end{figure}
In figure \ref{fig:can_H} we display the bundle contribution to the free energy $\beta \Omega^{bun}(L,t,\hrho,N_f)~+~N_f\ln\frac{q_20\Lambda^6}{K_0^2}=-\sum_{n=1}^{N_f} \ln\mathcal{D}(\hat{\rho}_1,L_n)$ for a bundle of $N_f=16$ filaments. We report results for both homogeneous and in-registry bundles for actin at the given room temperature persistence length, and for two values of the free monomers reduced density $\hrho$. We observe that for the specific d-WLC model of living filaments with force law given by Eq.(\ref{f_j}) the bundle contribution to the free energy is roughly linear with $L$ up to a $\hrho$-dependent value of the box size $L$ above which the presence of escaping filaments drives the system towards a different, unjustified, linear behavior. In the main figure results for both type of bundles, homogeneous and in-registry, appear to be superposed. However at the magnified scale of the inset an almost discontinuous behavior with period $d$ is seen for the in-registry case. At an even finer scale (not shown), the same behavior can be detected for the homogeneous bundle, although with a period of $d/N_f$. The nature and the origin of this behavior directly relies on the equilibrium force expression Eq.(\ref{Fbun}) \cite{Paper1}. An infinitesimal change of $L$, by a fraction of $d$, is accompanied by a strong variations of the $\alpha(j_n,L_n)$ factors for filament lengths touching the wall, and hence by a large modification of the equilibrium size distribution $\mathcal{P}(j_n|L_n,\hrho)$ and of the strength of the mean force $\bar{f}_j(L)$ exerted on the wall by a filament with size $j$. The dashed lines in the main panel of Fig.\ref{fig:can_H} correspond to the linear behavior $- F_{s}^{H} L$ based on Hill's mean field prediction of Eq.(\ref{F_hill}). This shows a close similarity of Hill's result with our free energy $\Omega^{bun}(L)$, up to the $\hrho$-dependent crossover to the escaping regime.
To avoid escaping, for given values of the parameters $\ell_p$ and $\hrho$, we have to impose a maximum size of the box; basing on the criterium given by Eq.(\ref{rho1lim}), we have to choose a length $L< L_{max}(\hrho,\ell_p)<\sqrt{\ell_pd/\ln\hrho}$. A good choice for $L_{max}$ follows imposing in Eq.(\ref{P_ratio}) a ratio of probabilities at most equal to 0.001,
\beq
\alpha(z^\ast(L_{max}),L_{max})\hat\rho_1^{\ z^\ast(L_{max})-z(L_{max})}=10^{-3}
\eeq
 giving $L_{max}=89.8d$ and $70.2d$ at $\hrho=1.7$ and 2.5 respectively.

In the optical trap apparatus the box size is a random variable which, by using Eqs.(\ref{Omega_fin}, \ref{eq:P(L)}) and the linearity shown in figure \ref{fig:can_H}, results to be Gaussian with a variance give by $\sigma_L=\sqrt{k_BT/\kappa_T}$.
Therefore a safe choice for the average box size of the optical trap has to be
\beq
\label{Lmax_OT}
\<L\>^{OT}<L_{max}(\hat\rho_1)-3\sqrt{\frac{k_BT}{\kappa_T }}<\sqrt{\frac{\ell_pd}{\ln\hrho}}-3\sqrt{\frac{k_BT}{\kappa_T}}
\eeq 
Knowing $\<L\>^{OT}$, Eq.(\ref{Lmax_OT}) provides a condition for the minimum value of the 
$\kappa_T $ that can be used for given $N_f$ and $\hrho$. A weaker trap would let the filaments become too long and eventually escape.
This is not the full story. To produce useful work, actin filaments in the usual conditions should not become shorter than $L_min=70~nm$ which corresponds to a minimum number of monomers $\sim 25$, as discussed by Mogilner \cite{Mogilner3}. Thus another constraint for the average optical trap size is: 
\beq
\<L\>^{OT}>L_{min}+3\sqrt{k_BT/\kappa_T }
\eeq
This additional constraint implies $\kappa_T< \kappa_{T,max}$.

\section{Results}
\label{sec5}
In this section we first compare rigid and flexible models for a single chain and a bundle of 8 filaments, the typical number in the experiments of ref. \cite{Footer}. Later we will investigate more in details flexible bundles for various number of filaments and for various average trap amplitudes. We stick on a single value of the reduced density, $\hrho=2.5$, a typical value for \textit{in-vitro} experiments \cite{Footer,Demoulin}. 
A more complete characterization of our F--actin model in different conditions is provided in the Supplementary Material.

\subsection{Rigid vs flexible behavior for single and 8-bundle filaments}
\begin{figure*}
\includegraphics[totalheight=10cm]{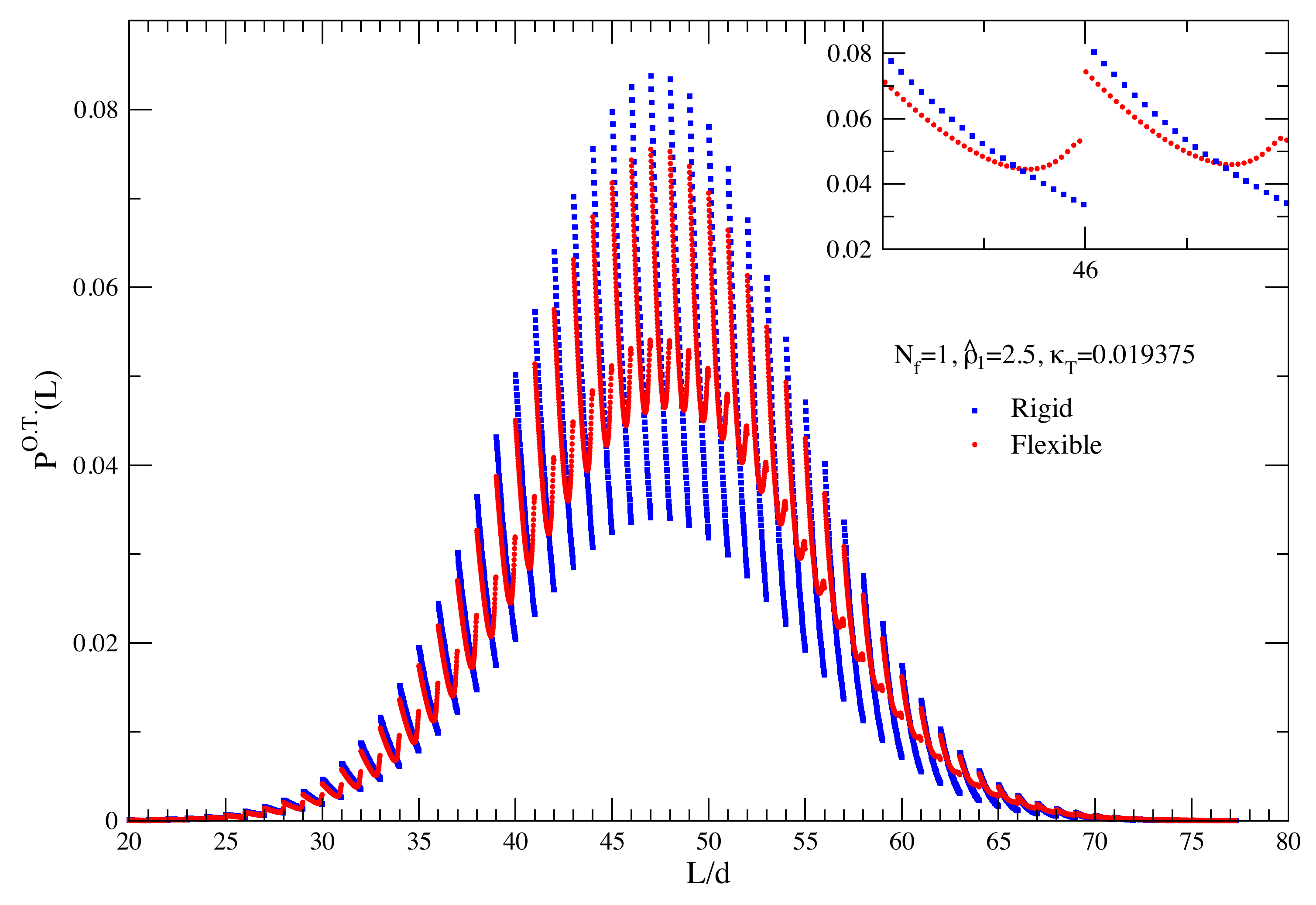}%width=0.8\columnwidth
\caption{Equilibrium distribution of the wall position $P^{O.T.}(L)$ for a single filament with $\hrho=2.5$ and $\kappa_T=0.019375$. Results for rigid (blue open squares) and flexible (red closed circles) cases are compared. Inset: detail of the distributions in the range $L\in[45;47]$ to show the reduction in discontinuity induced by the flexibility.}
\label{fig:single}
\end{figure*}

In figure \ref{fig:single} we show $P^{O.T.}(L)$, as computed by Eq.(\ref{eq:P(L)}), for rigid and flexible single filaments in a trap with $\kappa_T =0.019375$. 

For the flexible case, $P^{O.T.}(L)$ and $\<L\>=47.58$ %47.57600823
are obtained using a numerical integration scheme. The average value of the optical trap size can be compared to $L_H=\frac{N_f~k_BT}{d \kappa_T} \ln(\hrho)=47.29$ computed from the Hill's expression for the stalling force divided by $\kappa_T$. In the flexible case the average sizes results slightly larger.

For the rigid case ($\alpha$ either zero or one), the expressions for $P^{O.T.}(L)$ and $\<L\>^{O.T.}$ can be derived analytically (see the Appendix). One finds
\begin{equation}
P^{O.T.}(L)=\sqrt{\frac{2 \kappa_T }{\pi k_B T}}
\frac{\left(\hrho^{|\frac Ld|}-1\right) \e^{-\frac{L^2}{2\sigma^2d^2}}}%{2 k_B T}}}
{\sum_{i=2}^{\infty} \left(\hrho^i-1\right) \left[ {\rm erfc}\left(\frac{i}{\sqrt{2}\sigma}\right)-{\rm erfc}\left(\frac{i+1}{\sqrt{2}\sigma}\right)\right]}
\label{P^O.T.}
\end{equation}
where $|x|$ indicates the integer part of a real variable $x$, $\sigma=\left(d\sqrt{\beta\kappa_T}\right)^{-1}$ and  
\begin{equation}
\<L\>^{O.T.}=~\sqrt{\frac{2k_BT}{\pi\kappa_T }}  \frac{\sum_{i=2}^{\infty} \left(\hrho^i-1\right)~\left[\e^{-\frac{i^2}{2\sigma^2}}-\e^{-\frac{(i+1)^2}{2\sigma^2}}\right]}{\sum_{i=2}^{\infty} \left(\hrho^i-1\right) \left[ {\rm erfc}\left(\frac{i}{\sqrt{2}\sigma}\right)-{\rm erfc}\left(\frac{i+1}{\sqrt{2}\sigma}\right)\right]}
\label{eq:lav_rigid}
\end{equation}
giving $\<L\>^{O.T.}=47.29$, which is in agreement with Hill's prediction up to the 9th decimal place. 
By computing Eq.(\ref{eq:lav_rigid}) for decreasing values of $\kappa_T$ one can show that for a given supercritical parameter, the statistical mechanics average (\ref{eq:lav_rigid}) tends to Hill's value $L_H$ when the length of the filament goes to infinity. We indeed observe that $\<L\>^{O.T.}/L_H\to1$ exponentially fast when $\kappa_T\to0$, hence the trap gets infinitely large.

\begin{figure*}
\includegraphics[totalheight=10cm]{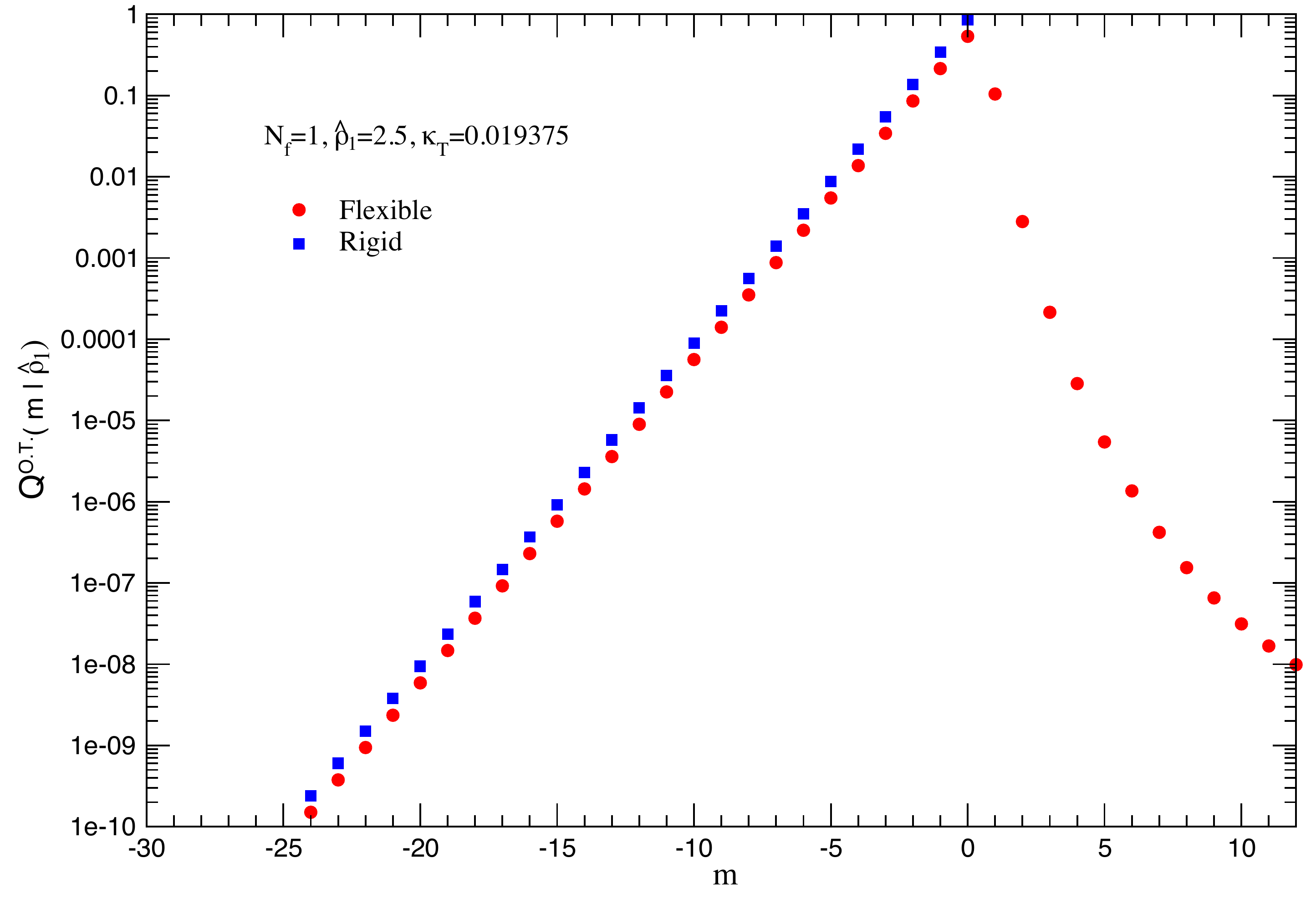}%width=0.8\columnwidth
\caption{Single filament relative size distribution $Q^{O.T.}(m|\hrho)$, Eq.(\ref{eq:Qm}), at  $\hrho=2.5$ in an optical trap with $\kappa_T =0.019375$. Results for rigid (blue open squares) and flexible (red closed circles) cases.}
\label{fig:Qm}
\end{figure*}
As for the probability of the single filament length with respect to the obstacle position, as defined in section \ref{sec2}, $Q^{O.T.}(m_n|\hrho)$, in figure \ref{fig:Qm} we report it for both rigid and flexible cases.
For the flexible case Eq.(\ref{eq:x0}) gives $\<x_0\>^{O.T.}=0.05528$, i.e. only roughly 6\% of the permitted filament lengths touch the wall during the brownian fluctuations of the wall inside the trap.

\begin{figure*}
\includegraphics[totalheight=8cm]{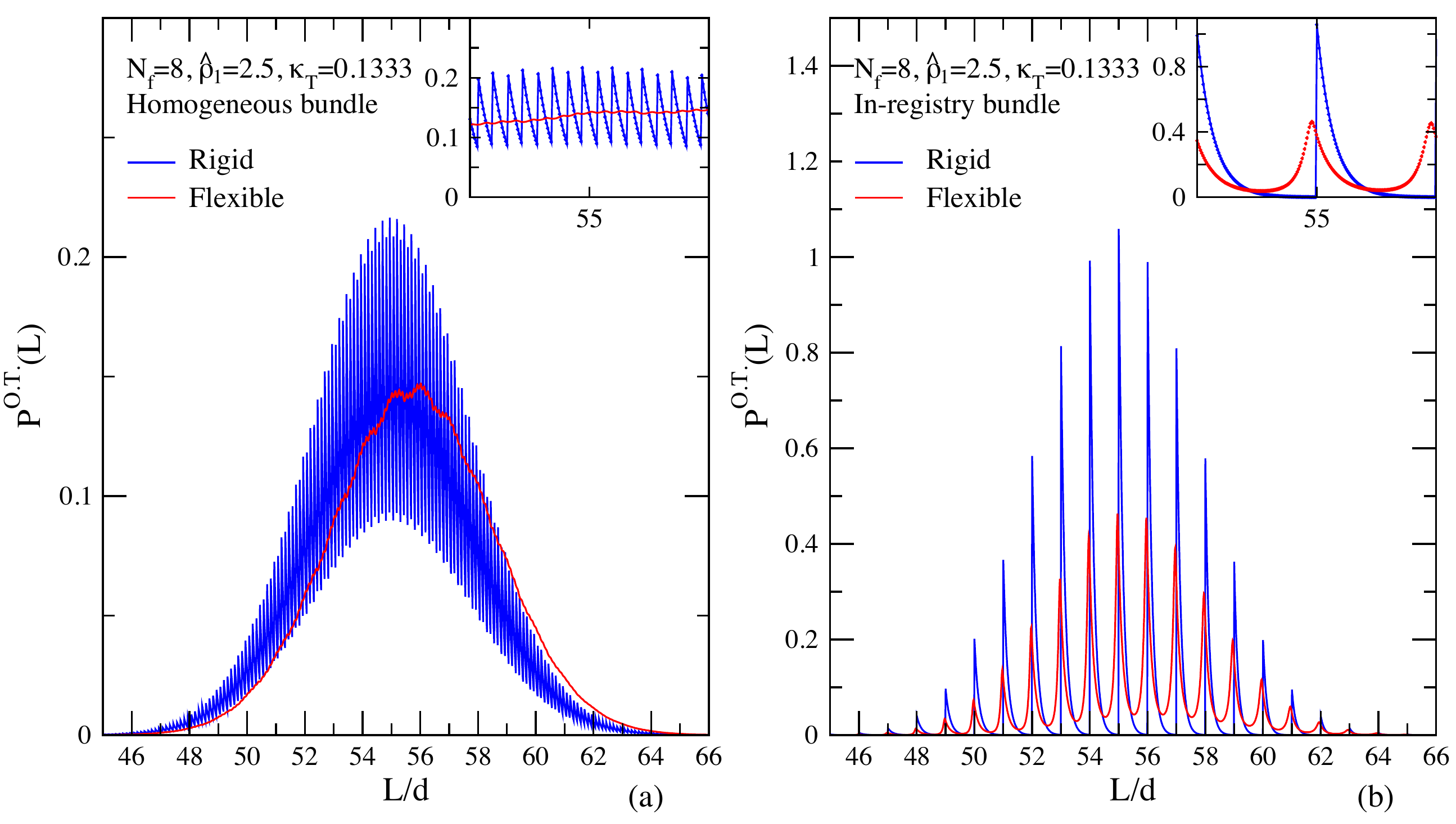}%width=0.8\columnwidth
\caption{$P^{O.T.}(L)$ for various 8-bundles at $\hrho=2.5$ in a trap with $\kappa_T=0.1333$. Panel (a): rigid (blue line) and flexible (red line) homogeneous bundles; panel (b): rigid (blue line) and flexible (red line) in--registry bundles. In both panels details of the distribution in a very limited $L$ range are given in the insets.}
\label{fig:Nf8}
\end{figure*}

For the bundle of $N_f=8$ filaments, figure \ref{fig:Nf8} compares $P^{O.T.}(L)$'s for flexible and rigid models of homogenous bundles (panel (a)) and in--registry bundles (panel (b)). Here we choose a trap strength $\kappa_T=0.1333$ which corresponds roughly to the upper $L$--limit of the non--escaping regime where flexibility effects are larger. The analytical expressions for $P^{O.T.}(L)$ and $\<L\>^{O.T.}$ for the rigid bundles are derived in the Appendix.
In the homogenous case we observe again an overall bell shape for both rigid and flexible bundles but the rigid case remains discontinuous (with a distance between successive jumps of $\Delta L/d=1/8$ now) while in the flexible case the $P^{O.T.}(L)$ becomes continuous although with some local oscillations arising from the strong rigidity of the single filaments (see the inset of panel (a)). Moreover $P^{O.T.}(L)$ of the flexible model is slightly shifted towards larger $L$ values since flexibility enhances the  bundle force \cite{Paper1} hence producing a larger average position of the trap. We obtain $\<L\>^{O.T.}=54.99$ for the rigid model again in perfect agreement with Hill's prediction (54.99), and $\<L\>^{O.T.}=55.75$ for the flexible model.
$P^{O.T.}(L)$ for in--registry bundles exhibits much stronger features. In the rigid case the discontinuities observed for the single filaments are strongly enhanced providing a series of nearly isolated peaks with the maximum at integer values of $L/d$. The amplitude of the minima are between 10$^{-6}$ at the tails of the distribution and 10$^{-4}$ at $L\sim 55$ which implies the presence of rather large free energy barriers in moving $L$ from one probability maximum to the next. The flexible case exhibits again an overall bell shape with local maxima at the same locations than for the rigid case but the behavior between successive peaks is continuous (see the inset in panel (b)). Values for the average trap lengths are $\<L\>^{O.T.}=55.57$ and $\<L\>^{O.T.}=54.99$ for flexible and rigid case respectively, the latter again in perfect agreement with Hill's prediction. As for the flexible case, we note that the trap length for the in--registry bundle is slightly smaller than for the homogenous bundle which reflects an effective larger stiffness of the in--registry disposition with respect to the homogenous disposition of seeds.

\subsection{Flexible $N_f$--bundles and mechanism of bundle force generation}
Below we discuss the behavior of flexible bundles in various regimes and we illustrate the mechanism used by the bundle to generate the force resisting the external load. In order to study the effect of $N_f$ at the same physical conditions, we compare results for different $N_f$ at the same $\<L\>^{O.T.}\approx L_H=\frac{N_f~k_BT}{d \kappa_T}\ln(\hrho)$. This requires to increase $\kappa_T$ linearly with $N_f$. In this way we can investigate the effect of flexibility.
\begin{figure*}
\includegraphics[width=0.8\columnwidth]{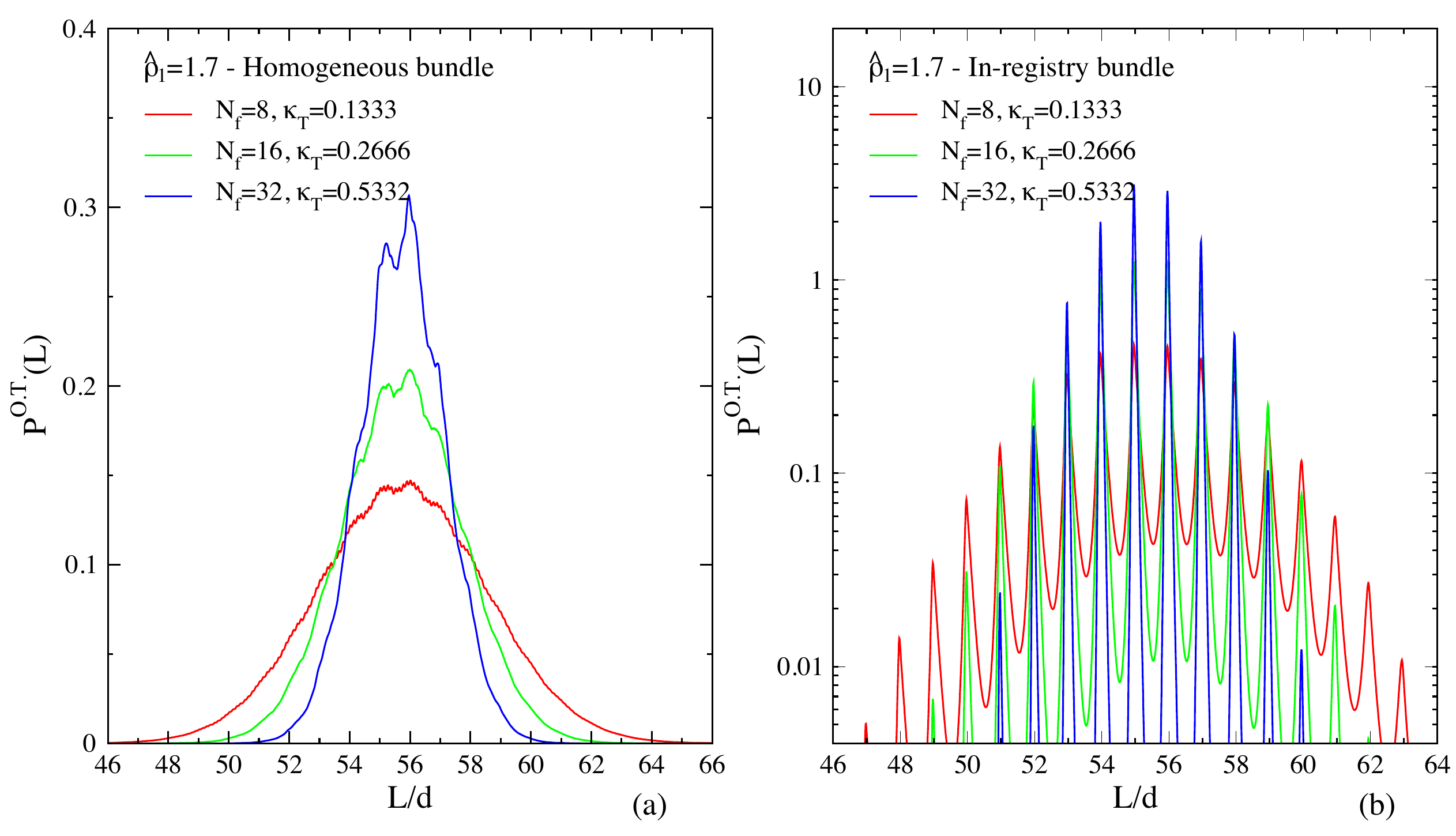}
\caption{$P^{O.T.}(L)$ for flexible bundles of $N_f=8$ (red curve), 16 (green curve) and 32 (blue curve) filaments at $\hrho=2.5$ and $\kappa_T=0.1333, 0.2666, 0.5332$ respectively. Panel (a): homogeneous bundles, panel (b): in--registry bundles.}
\label{fig:P(L)_flexible}
\end{figure*}
In figure \ref{fig:P(L)_flexible} we show $P^{O.T.}(L)$ for flexible bundles of $N_f=8, 16, 32$ filaments at the same value $L_H=55.74$ (we checked that we are in the non-escaping regime).

For homogenous bundles the shape of $P^{O.T.}(L)$ is well represented by a gaussian function centred at $\<L\>^{O.T.}$ with a width decreasing as $\kappa_T^{-1/2}\propto N_f^{-1/2}$ and some additional features around the maximum increasing with $N_f$. The average trap length $\<L\>^{O.T.}$ is essentially independent of $N_f$. For large $\<L\>^{O.T.}$ like the ones in figure \ref{fig:P(L)_flexible}, the relative deviation of $\<L\>^{O.T.}$ from $L_H$ is 1.4\% and decreases with decreasing flexibility. 

$P^{O.T.}(L)$ for in--registry bundles are presented in panel (b) of figure \ref{fig:P(L)_flexible}. The distributions are very different from the corresponding homogeneous case since now they have very strong oscillations superimposed to the gaussian behavior. However, as for the homogeneous case, the distribution gets more localized for increasing $N_f$. The relative deviation of $\<L\>^{O.T.}$ from $L_H$ at given $N_f$ is smaller than in the corresponding homogenous case.

\begin{figure*}
\centering
% \subfigure[]
%   {\includegraphics[width=9.5cm]{NEW_FIG/P_F.pdf}}
% \hspace{2mm}
% \subfigure[]
%   {\includegraphics[width=8cm]{NEW_FIG/NF16_exp5LFdist_2.pdf}}
\includegraphics[width=0.9\columnwidth]{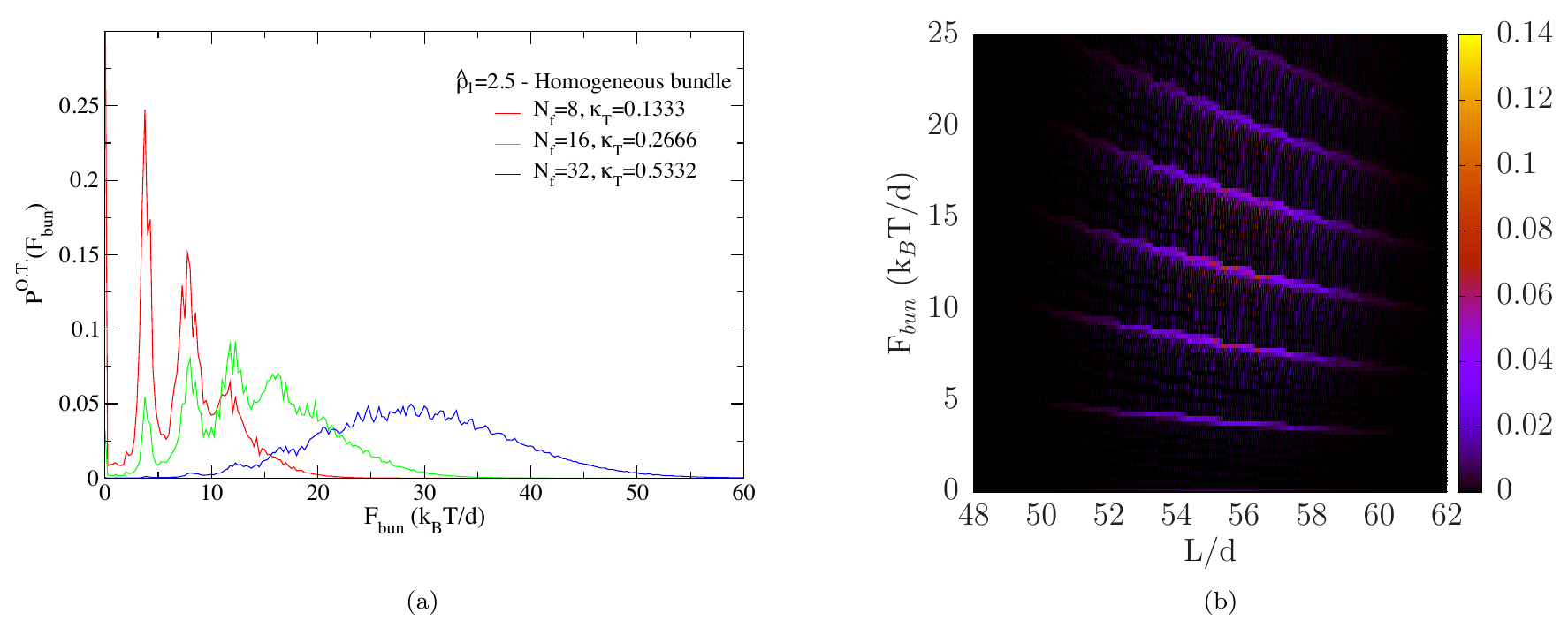}
\caption{Panel (a): $P^{O.T.}(F_{bun})$ for the same systems shown in figure \ref{fig:P(L)_flexible}. Panel (b): colormap of $P^{O.T.}(L,F_{bun})$ for the 16--bundle.}
\label{fig:P(F)_flexible}
\end{figure*}

As for the bundle force and its dependence on $N_f$ at given $L_H$, we concentrate here on the homogenous bundles of $N_f=8, 16$ and 32 filaments. We report in figure \ref{fig:P(F)_flexible} the distribution of the bundle force (panel (a)) defined as:
\begin{equation}
P^{O.T.}(F_{bun})\equiv\Big<\delta\left(F_{bun}-\sum_{n=1}^{N_f}\bar{f}_{j_n}(L_n)\right)\Big\>=\int_0^{L_R}dL\sum_{\{j_n\}}\Bigg\{\delta\left(F_{bun}-\sum_{n=1}^{N_f}\bar{f}_{j_n}(L_n)\right)\left[\prod_{n=1}^{N_f}\mathcal{P}(j_n|L_n,\hrho)\right]P(L)\Bigg\}
\label{eq:P(F)}
\end{equation}
where $\sum_{n=1}^{N_f}\bar{f}_{j_n}(L_n)$ is the sum of the force exerted by each filament when the configuration of the system is $\{j_1,\dots,j_n,L\}$, given by Eq(\ref{f_j}).
For all bundles $P^{O.T.}(F_{bun})$ has peaks at specific values of the force in the low force range and a roughly gaussian overall behavior. The amplitude of the peaks decreases strongly with $N_f$: they are barely visible for $N_f=32$.
Despite the peculiar differences of $P^{O.T.}(F_{bun})$ for the three bundles, the average bundle force per filament does not depend on $N_f$ and is $\<F_{bun}\>^{O.T.}/N_f= 0.9289$ to be compared to Hill's value $F_s^H=0.9163$, again a genuine effect of flexibility since fully rigid bundles provide results in perfect agreement with Hill's theory. 
The position of the maximum of the gaussian envelope and the average force values are extensive with $N_f$ while the position of the peaks at small force values does not depend on $N_f$ or on the disposition of the filament seeds (in--registry bundles present the same peaks) which indicates that the single filaments are responsible for this behavior. The distance between two adjacent peaks is roughly equal to the value of the ``buckling" force of the individual filaments  of contour length $L_c\approx \langle L\rangle^{O.T.}$, $f_b(L_c,\ell_p)$, as defined in Eq.(\ref{eq:fb_f||}). Indeed for the investigated width of the trap $\langle L\rangle^{O.T.}\approx 55.7 d$, $f_b=\frac{\pi^2 \ell_p k_BT}{4 L_c^2}\approx  4.2~k_BT/d$ in reasonable agreement with the observation in panel (a) of figure \ref{fig:P(F)_flexible}. 

To better understand the nature of the observed bundle force distributions we have investigated the joint probability $P^{O.T.}(L,F_{bun})$, defined as
\begin{equation}
P^{O.T.}(L,F_{bun})=\sum_{\{j_n\}}\delta\left(F_{bun}-\sum_{n=1}^{N_f}\bar{f}_{j_n}(L_n)\right)\prod_{n=1}^{N_f}\mathcal{P}(j_n|L_n,\hrho)P(L)
\label{eq:P(L,F)}
\end{equation}
These data are obtained by computing the expected value over the $\{j_n\}$ through a Monte Carlo sampling of the variables $L$ and then $\{j_n\}$. We see how the structure of the marginal $P^{O.T.}(F_{bun})$ comes from the very specific form of the joint probability distribution: the joint distribution exhibits a sequence of crests and valleys which extend along the $L$ direction and are related to the ``discrete'' character of the force exerted by the individual buckled filaments. 
Moving along each crest from small to large $L$ values, a maximum of probability is present around $L=55$.
In the small force range the crests are almost parallel to the $L$ axis which results in the marked peaks observed in the marginal distribution $P^{O.T.}(F_{bun})$, while in the large force range the crests get progressively tilted by a negative angle with respect to the $L$ axis which produces the continuous tail observed in $P^{O.T.}(F_{bun})$. Along the individual crests, we see that the force decreases for increasing $L$ which is in agreement with the buckling force expression. It is reasonable to assign each crest to a specific number of buckled filaments participating to the average bundle force: in panel (b) the first crest (from lower force) is due to the a single filament pushing the obstacle, the second crest is due to two filaments, and so on. 
Therefore the peaks observed in $P^{O.T.}(F_{bun})$ arise from the contribution to the bundle force of integer numbers of buckled filaments. Note that filaments in contact with the obstacle are mostly buckled since the buckling regime of the compression force starts at quite small reduced compression ($\tilde{\eta} \simeq 0.25$). Once a filament is buckled it will oppose a force independent of the applied compression. However, due to its living character, it can change its contour length and exit the buckling regime by loosing one or few monomers. At the same time, filaments whose tip position is closer to the wall position by less than a monomer size $d$ can increase their contour length by polymerization and, if the filaments stiffness is large enough (which means for shorter filaments), it can enter in the buckled state. 
We can look at this mechanism by a two--state model: filaments are either non--active (not in contact with the obstacle and therefore providing zero contribution to the bundle force) or in their buckled state therefore contributing to the force by a finite amount equal to $f_b(L_c)\propto L_c^{-2}\approx L^{-2}$. In this two-states model a living bundle of given length can adjust the number of touching filaments to resist to an external load. 

In the present optical trap system the equilibrium bundle force can be related to the average number of touching filaments as follows 
\bea
&&\<F_{bun}\>^{O.T.}=\int_0^{L_R}dLP^{O.T.}(L)F_{bun}(L)\nonumber\\
&=&\int_0^{L_R}dLP^{O.T.}(L)\sum_{n=1}^{N_f}\sum_{j_n=z_n}^{z_n^\ast}\frac{\ \pi^2}4\frac{k_BT\ell_p}{L_{c,j}^2}\tilde{f}_{\parallel}(\tilde\eta_{j_n})\mathcal{P}(j_n|L_n)\nonumber\\
&\approx&\frac{\ \pi^2}4\frac{k_BT\ell_p}{\(\<L\>^{O.T.}\)^2}\int_0^{L_R}dLP^{O.T.}(L)\sum_{n=1}^{N_f}\sum_{j_n=z_n}^{z_n^\ast}\mathcal{P}(j_n|L_n)\nonumber\\
&=&\frac{\ \pi^2}4 \frac{k_BT\ell_p}{\(\<L\>^{O.T.}\)^2}\<N_0\>^{O.T.}
\eea
where we used Eq.(\ref{eq:n0}) and the approximations $L_{c,j}^2\approx(\<L\>^{O.T.})^2$ and $\tilde{f}_{\parallel}(\tilde\eta_{j_n})=1$. 
As the average bundle force roughly equals Hill's stalling force $F_{s}^{H}=N_f(k_BT/d) \ln\hrho$, one gets
\beq
\label{eq:x0_L2}
\frac{\<x_0\>^{O.T.}}{\ln\hat\rho_1}\approx\frac4{\ \pi^2} \frac{1}{\ell_pd}\(\<L\>^{O.T.}\)^2
\eeq
having defined $\<N_0\>^{O.T.}=\<x\>^{O.T.}N_f$.
\begin{figure}[!]
\includegraphics[width=0.6\columnwidth]{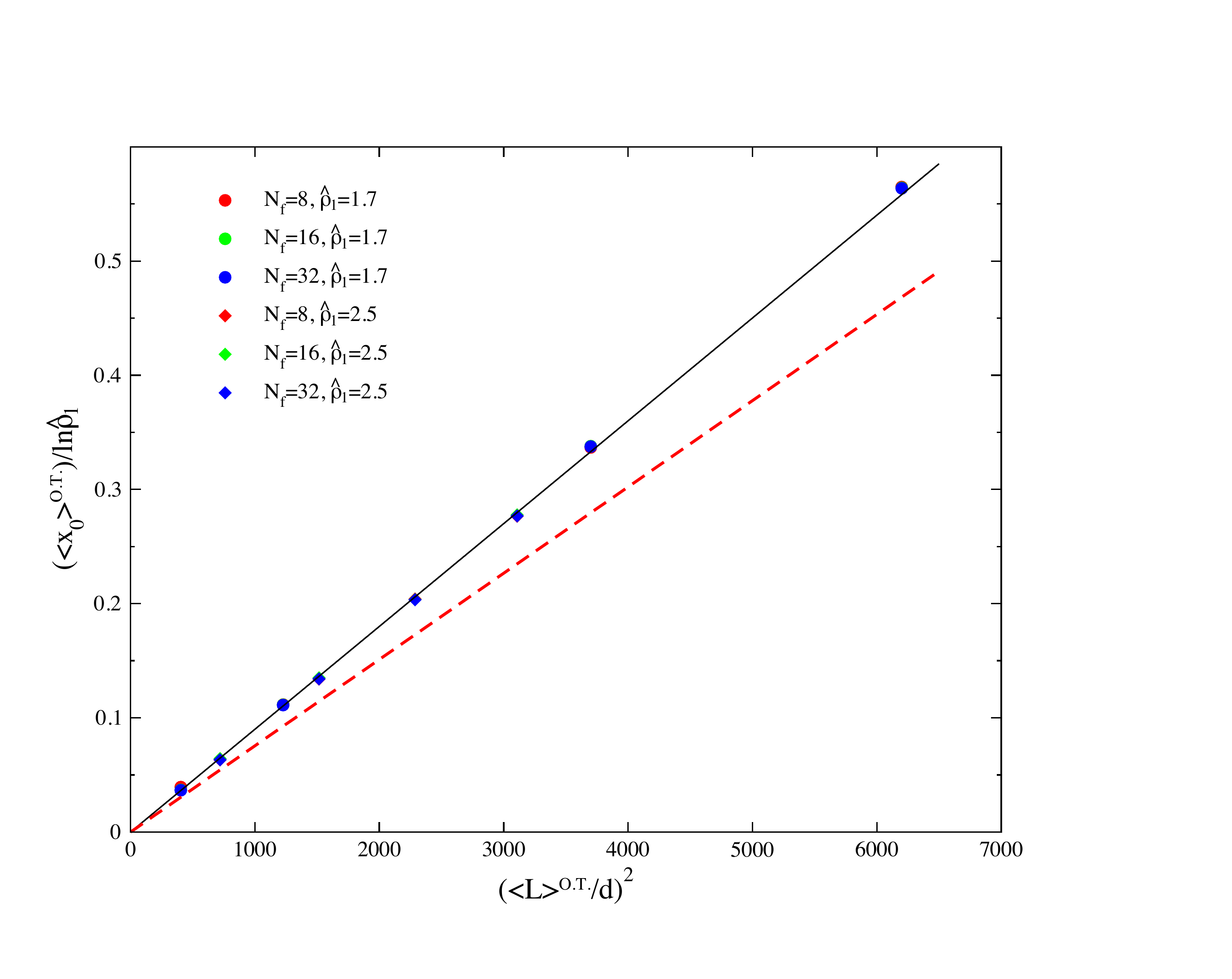}
\caption{Average fraction of touching filaments, divided by $\ln\hrho$, as a function of the average wall position squared. Data points for homogeneous bundles with different $N_f$ and two values of $\hrho$ collapse on a linear behavior $\<x_0\>^{O.T.}/\ln\hrho \propto(\<L\>^{O.T.}/d)^2$, but with a 15-20$\%$ disagreement with the slope given by Eq.(\ref{eq:x0_L2}) (red dashed line).}
\label{fig:x0L2}
\end{figure}
Figure \ref{fig:x0L2} shows the average fraction of touching filaments in the homogeneous bundle, divided by $\ln(\hrho)$,
as a function of $\(\<L\>^{O.T.}\)^2$ for two different 
densities  and various number of filaments.
Data points align on a line with slope $\sim9\times10^{-5}$, represented by the straight black line in Figure \ref{fig:x0L2}. The red dashed line corresponds to the slope expected according to Eq.(\ref{eq:x0_L2}), slightly smaller ($\approx7.5\times10^{-5}$): the number of touching filaments results slightly larger than predicted by Hill's theory. 
This is a further effect of filament flexibility. Since the average polymerization force is slightly larger than Hill's stalling force, Eq.(\ref{eq:x0_L2}) underestimates the observed average fraction of touching filaments.

\section{Discussion and Final remarks} \label{sec6}
Before concluding, it is interesting to discuss the experimental situation of ref.\cite{Footer}: the growth of $N_f\approx8$ filaments bundle anchored to a latex bead controlled by a harmonic force from an optical trap apparatus, and pushing against an immobile and impenetrable surface was followed up to the establishing of a stationary state (see figure 4 of reference \cite{Footer}). Two different sets of experiments at different free monomers density, $\rho_1=4$ and $2~\mu M$ were performed, corresponding in our chemically simplified model to $\hrho=2.5$ and $\hrho=1.7$, respectively. The measured latex bead displacement  at stationarity was $\<L\>\approx900 nm\approx300d$ at $\hrho=2.5$ and to $\<L\>\approx180 nm\approx70d$ at $\hrho=1.7$. Knowing the strength of the trap apparatus, $0.0035~k_BT/d^2$ in the first case and $0.011~k_BT/d^2$ in the second case, the measured displacements corresponded to an apparent stalling force of $\<F_{bun}\>^{exp}\approx1.6~pN=1.05~k_BT/d$ at $\hrho=2.5$ and $1.1~pN=0.72~k_BT/d$ at $\hrho=1.7$. Those values are in marked disagreement with Hill's formula, Eq.(\ref{F_hill}), which predicts $F_s^H=6.31 k_BT/d$ and $4.25~k_BT/d$ respectively.
The analysis of these experiments was based on the distinction between buckled and unbuckled filaments, using the following criterium: if the detected force was smaller than the single filament buckling force, Eq.(\ref{eq:fb_f||}), at the observed average length of the bundle, then the measured force was considered a proper estimate of the stalling force, otherwise the measured force was regarded as meaningless. The first case ($\hrho=2.5$) fell within this latter condition, and hence it was not considered. The second case ($\hrho=1.7$) instead fell into the former condition and therefore was considered a correct measure of the bundle polymerization force.
According to our analysis of section \ref{sec4_b} which assumes the absence of hydrolysis, we find that the average trap length should be below $L_{max}-3\sigma_L$ to avoid the occurrence of escaping filaments (see Eq.(\ref{Lmax_OT})). At $\hrho=2.5$ the measured steady bead position in the experiment was much larger than $L_{max}$, so we can conclude that the bundle was in the escaping regime, in agreement with the interpretation of the authors. In the other case the observed average trap length did not exceed $L_{max}$, but it still exceeded the boundary given by Eq.(\ref{Lmax_OT}), as $\<L\>^{O.T.}\sim70d>90d-3(0.011)^{-1/2}d\simeq60d$ so that we should expect the presence of escaped filaments. The detected force of $\<F_{bun}\>^{exp}\simeq 0.72~k_BT/d$ is about $35\%$ larger than the stalling force of a single filaments $k_BT\ln\hrho/d$. It is possible that the residual force accounts for the elastic force from seven escaped filaments.
According to our analysis, what is needed to measure the polymerization force of a 8-bundle at $\hrho=1.7$ is a stronger trap with $\kappa_T \approx0.1~k_BT/d^2$, establishing the equilibrium distance around $40\div50d$ well within the non-escaping regime.

In conclusion, in this paper we have developed the Statistical Mechanics formalism to treat a bundle of (de)-polymerizing filaments in a box pushing against a mobile wall. Our system is a schematic representation of \textit{in--vitro} experimental apparatus exploited to measure the force that a bundle of F-actin can exert on an obstacle. Our treatment, limited to equilibrium conditions, requires the external load to increase with the distance between the channel boundaries in order to match the bundle force at stalling, where the bundle growth is stopped and a genuine equilibrium state is established. 
We have developed the formalism for a simple and flexible model in which we have disregarded direct inter-filaments and filament-solvent interactions which are considered to be irrelevant in the present context. The formalism has been used for two specific filament models under a load increasing linearly with the box size: i) the fully rigid model (1D) which is at the heart of the much celebrated Brownian Ratchet model used in interpreting experimental data for the force-velocity law \cite{Footer,Demoulin} and ii) a model of semiflexible discrete Wormlike chain with persistence length $\ell_p$ and monomer size $d$ adapted to F-actin values for which a force-compression law is known from previous studies \cite{Frey,Paper1}. For the rigid model, we have derived exact expressions for the probability distribution of the mobile obstacle position and its average value, for a single filament and for homogeneous and in--registry bundles. These expressions allow us to discuss the validity of the celebrated Hill's formula for the stalling force. We found that for box sizes beyond $\sim 5d$ our exact statistical mechanics averages, taken over the optical trap ensemble, do converge asymptotically and exponentially fast to the Hill's prediction based on 1D thermodynamic approach with the relative deviation being already $\leq 1\%$ at $\<L\>\sim 10d$ and $\sim 10^{-6}$ for $\<L\>\sim 30d$. For narrower boxes ($\<L\>< 10d$), exact results indicate a markedly different behavior, a boundary effect already noticed in ref. \cite{Paper1}.
The consideration of filament flexibility forces us to distinguish the \textit{stalling regime}, the regime of small optical trap widths in which the external load is able to stop the polymerization of all filaments, from the \textit{escaping regime}, the regime of larger trap widths where the bent filaments can polymerize freely parallel the obstacle wall and the external load is balanced by the mechanical bending force of the filaments. In the stalling regime we found that, with respect to the rigid filament case, flexibility induces in general a slight increase of the equilibrium optical trap distance and of the stalling force, the amount of which depends on the average optical trap width and on the disposition of the bundle seeds. A marked difference from the rigid model is that the individual filaments are either not in contact with the obstacle, hence not directly active, or in their buckled state providing a finite force roughly proportional to $(\<L\>^{O.T.})^{-2}$. Therefore the total bundle force results, to a very good level of accuracy, from $N_f$ times the average fraction $\<x_0\>$ of touching filaments each one resisting with its buckling force. The observation of a bundle force largely independent of the trap width results in a fraction of touching filaments increasing with $(\<L\>^{O.T.})^2$.

\section*{Acknowledgements}
Two of us (CP and JPR) would like to thank M. Baus, M. J. Footer, and B. Mognetti for useful discussions. We thank G. Destree and P. Pirotte for technical help. This work has been supported by the Italian Institute of Technology (IIT) under the SEED project Grant No. 259 SIMBEDD and by the Italian Ministry of Research under Project No. PRIN2012--2012NNRKAF.

\appendix
\section{Rigid filament model}
In this appendix we derive the analytical expressions for $P^{O.T.}(L)$ and $\<L\>^{O.T.}$ for the rigid model of single filaments and homogeneous and in--registry bundles.

\subsection{Single filament}
For a single rigid filament ($\alpha(j,L)$ either one or zero) at given $\hrho$ and in a box of fixed size $L$, the filament partition function $\mathcal{D}(L,\hat\rho_1)$, as defined in Eq.(\ref{Drig}), is
\beq
\mathcal{D}(L,\hat\rho_1)=\begin{cases}\sum_{j=2}^{z^\ast}\alpha_j(L)\hrho^j=
\frac{\hrho^2}{\hrho-1} \left(\hrho^{|\frac{L}{d}|}-1 \right)\qquad &\mbox{$2d<L<\infty$}\\
0 &\mbox{otherwise}
\end{cases}
\label{eq:D_single}
\eeq
where $|\cdots|$ is the integer part of the argument.
The normalization of the trap size distribution in the optical trap ensemble is 
\beq
N=\int_0^{\infty} dL~ \mathcal{D}(L,\hat\rho_1)~\e^{-\frac{\beta \kappa_t L^2}{2}}
\eeq
where the upper limit of the integral has been taken to $\infty$ since we are considering rigid filaments. Changing the integration variable, in the range of continuity of the step-shaped integrand, to $y=L/(\sqrt{2}d\sigma)$ with $\sigma^2=(\beta \kappa_T d^2)^{-1}$, we obtain
\bea
N&=&\frac{\hrho^2 d}{\hrho-1} \sum_{i=2}^{\infty} \left(\hrho^i-1\right) \sqrt{2}\sigma \int_{i/\sqrt{2}\sigma}^{(i+1)/\sqrt{2}\sigma} dy~ \e^{-y^2} \nonumber \\
&=&\frac{\hrho^2 d}{\hrho-1}  \sqrt{\frac{\pi}{2}}\sigma \sum_{i=2}^{\infty} \left(\hrho^i-1\right) \left[ {\rm erfc}\left(\frac{i}{\sqrt{2}\sigma}\right)-{\rm erfc}\left(\frac{i+1}{\sqrt{2}\sigma}\right)\right]% \nonumber\\
%&=&\frac{\hrho^2 d}{\hrho-1}\sqrt{\frac{\pi}{2}}\sigma \left[\sum_{i=0}^{\infty} \left(\hrho^i-1\right){\rm erfc}\left(\frac{i}{\sqrt{2}\sigma}\right)-\sum_{i=1}^{\infty} \left(\hrho^{i-1}-1\right){\rm erfc}\left(\frac{i}{\sqrt{2}\sigma}\right) \right]\nonumber\\
%&=&\hrho~ d ~\sqrt{\frac{\pi}{2}}\sigma ~\sum_{i=1}^{\infty} \hrho^i~ {\rm erfc}\left(\frac{i}{\sqrt{2}\sigma}\right)
\eea
With this results we have 
\bea
P^{O.T.}(L)&=&N^{-1}~ \frac{\hrho^2}{\hrho-1} \left(\hrho^{|\frac{L}{d}|}-1 \right) \e^{-\frac{\kappa_T L^2}{2 k_B T}}\nonumber \\
%&=&\frac{\hrho}{\hrho-1}
&=&\sqrt{\frac{2  }{\pi}}\left(\sigma d\right)^{-1}
\frac{\left(\hrho^{|\frac Ld|}-1\right) \e^{-\frac{L^2}{2\sigma^2d^2}}}%{-\frac{\kappa_T L^2}{2 k_B T}}}
{\sum_{i=2}^{\infty} \left(\hrho^i-1\right) \left[ {\rm erfc}\left(\frac{i}{\sqrt{2}\sigma}\right)-{\rm erfc}\left(\frac{i+1}{\sqrt{2}\sigma}\right)\right]}
%{\sum_{i=1}^\infty \hrho^i ~{\rm erfc}\left(i\sqrt{\frac{\kappa_T  d^2}{2 k_B T}}\right)}
\eea
i.e. Eq.(\ref{P^O.T.}), \textit{qed}.

The average trap length $\<L\>^{O.T.}$ is ($x=L/d$ and $y=x^2$) 
\bea
\<L\>^{O.T.} &=& \int_0^{\infty} dL~ P^{O.T.}(L)~L=N^{-1}\frac{\hrho^2 d^2}{\hrho-1} \int_0^{\infty} dx~x~ \left(\hrho^{|x|}-1 \right) \e^{-\frac{x^2}{2\sigma^2}} \nonumber \\
&=& N^{-1}\frac{\hrho^2 d^2}{\hrho-1} \sum_{i=2}^{\infty} \left(\hrho^i-1\right)~ \frac{1}{2}~\int_{i^2}^{(i+1)^2} dy~ \e^{-\frac{y}{2\sigma^2}} =\nonumber\\%%%
&=&  N^{-1}\frac{\hrho^2 d^2\sigma^2}{\hrho-1} \sum_{i=2}^{\infty} \left(\hrho^i-1\right)~\left(\e^{-\frac{i^2}{2\sigma^2}}-\e^{-\frac{(i+1)^2}{2\sigma^2}}\right)\nonumber\\
&=&\sqrt{\frac{2}{\pi}}\left(\sigma d\right) \frac{\sum_{i=2}^{\infty} \left(\hrho^i-1\right)~\left[\e^{-\frac{i^2}{2\sigma^2}}-\e^{-\frac{(i+1)^2}{2\sigma^2}}\right]}{\sum_{i=2}^{\infty} \left(\hrho^i-1\right) \left[ {\rm erfc}\left(\frac{i}{\sqrt{2}\sigma}\right)-{\rm erfc}\left(\frac{i+1}{\sqrt{2}\sigma}\right)\right]}
%&=&  N^{-1}\frac{\hrho^2 d^2\sigma^2}{\hrho-1} \sum_{i=0}^{\infty} \left(\hrho^i-1\right)~\e^{-\frac{i^2}{2\sigma^2}}\left(1-\e^{-\frac{2i+1}{2\sigma^2}}\right)=\sqrt{\frac{2}{\pi}}~d~\sigma\frac{\sum_{i=1}^{\infty} \hrho^i~\e^{-\frac{i^2}{2\sigma^2}}}{\sum_{i=1}^{\infty} \hrho^i~ {\rm erfc}\left(\frac{i}{\sqrt{2}\sigma}\right)}\\
%&=&\sqrt{\frac{2}{\pi}}~\frac{k_BT}{\kappa_T d} \frac{\sum_{i=1}^{\infty} \hrho^i~e^{-\frac{i^2 \kappa_T d^2}{2 k_BT}}}{\sum_{i=1}^{\infty} \hrho^i~ {\rm erfc}\left(i\sqrt{\frac{\kappa_T  d^2}{2 k_B T}}\right)}
\eea
i.e. Eq.(\ref{eq:lav_rigid}). 

\subsection{In--registry bundle}
For an in--registry bundle of $N_f$ filaments the bundle partition function at fixed $L$ is given by the product of $N_f$ identical single filament partition functions, Eq.(\ref{eq:D_single}),
\beq
\mathcal{D}(L,\hat\rho_1)=\begin{cases}\left(\frac{\hrho^2}{\hrho-1}\right)^{N_f} \left(\hrho^{|\frac{L}{d}|}-1 \right)^{N_f}\qquad&\mbox{$2d<L<\infty$}\\
0 &\mbox{otherwise}
\end{cases}
\eeq
while the normalization of $P^{O.T.}(L)$ is 
\bea
N&=&\sqrt{\frac{\pi}{2}}~d~\sigma \left(\frac{\hrho^2}{\hrho-1}\right)^{N_f} \sum_{i=2}^{\infty} \left(\hrho^i-1 \right)^{N_f}\left[ {\rm erfc}\left(\frac{i}{\sqrt{2}\sigma}\right)-{\rm erfc}\left(\frac{i+1}{\sqrt{2}\sigma}\right) \right]%\nonumber \\
%&=&\sqrt{\frac{\pi}{2}}~d~\sigma \left(\frac{\hrho^2}{\hrho-1}\right)^{N_f} \sum_{i=1}^{\infty} \left[ \left(\hrho^{i}-1 \right)^{N_f}-\left(\hrho^{i-1}-1 \right)^{N_f}\right] {\rm erfc}\left(\frac{i}{\sqrt{2}\sigma}\right)
\eea
Therefore the wall position probability distribution in the optical trap is given by
\beq
P^{O.T.}(L)=\sqrt{\frac{2}{\pi}}\sqrt{\frac{\kappa_T}{k_BT}}\frac{\left(\hrho^{|\frac{L}{d}|}-1 \right)^{N_f} \e^{-\frac{(L/d)^2}{2\sigma^2}}}{\sum_{i=2}^{\infty} \left(\hrho^i-1 \right)^{N_f}\left[ {\rm erfc}\left(\frac{i}{\sqrt{2}\sigma}\right)-{\rm erfc}\left(\frac{i+1}{\sqrt{2}\sigma}\right) \right]}
%=\sqrt{\frac{2}{\pi\sigma^2}}\frac{\left(\hrho^{|\frac{L}{d}|}-1 \right)^{N_f} \e^{-\frac{(L/d)^2}{2\sigma^2}}}{d \sum_{i=1}^{\infty} \left[ \left(\hrho^{i}-1 \right)^{N_f}-\left(\hrho^{i-1}-1 \right)^{N_f}\right] {\rm erfc}\left(\frac{i}{\sqrt{2}\sigma}\right)}
\eeq
Following a procedure similar to that for the single filament, we obtain for the average trap length of this bundle
\bea
\<L\>^{O.T.} &=& \sqrt{\frac{2}{\pi}}~\sqrt{\frac{k_BT}{\kappa_T }}\frac{\sum_{i=2}^{\infty} \left(\hrho^i-1\right)^{N_f}~\left[\e^{-\frac{i^2}{2\sigma^2}}-\e^{-\frac{(i+1)^2}{2\sigma^2}}\right]}{\sum_{i=2}^{\infty} \left(\hrho^i-1 \right)^{N_f}\left[ {\rm erfc}\left(\frac{i}{\sqrt{2}\sigma}\right)-{\rm erfc}\left(\frac{i+1}{\sqrt{2}\sigma}\right) \right]}
%\sqrt{\frac{2}{\pi}}d~\sigma\frac{\sum_{i=0}^{\infty} \left(\hrho^i-1\right)^{N_f}~\e^{-\frac{i^2}{2\sigma^2}}\left(1-\e^{-\frac{2i+1}{2\sigma^2}}\right)}{\sum_{i=1}^{\infty} \left[ \left(\hrho^{i}-1 \right)^{N_f}-\left(\hrho^{i-1}-1 \right)^{N_f}\right] {\rm erfc}\left(\frac{i}{\sqrt{2}\sigma}\right)}
\eea 

\subsection{Homogenous bundle}
 According to the definition given in the main text, in a homogeneous bundle of $N_f$ filaments the filament $n$ starts at 
\beq
h^\ast_n\equiv\frac{h_n}d=\frac{n}{N_f}-\frac{1}{2N_f}-\frac{1}{2} \qquad\qquad\qquad\qquad n\in[1,N_f]
\eeq
The bundle partition function is now
\beq
\mathcal{D}(L,\hat\rho_1)=\begin{cases}\left(\frac{\hrho^2}{\hrho-1}\right)^{N_f} \prod_{n=1}^{N_f}\left(\hrho^{|L/d-h^\ast_n|}-1 \right)\qquad&\mbox{$2d-h_{N_f}<L<\infty$}\\
0&\mbox{otherwise}
\end{cases}
\eeq
and the normalization of $P^{O.T.}(L)$ 
\bea
N=\left(\frac{\hrho^2}{\hrho-1}\right)^{N_f} d \int_0^{\infty} dx~ \prod_{n=1}^{N_f}\left(\hrho^{|x-h^\ast_n|}-1 \right) \e^{-\frac{x^2}{2\sigma^2}}
\eea
with $x=L/d$. $h_{N_f}>h_{N_f-1}>\dots>h_1$ implies $x-h_{N_f}<x-h_{N_f-1}<\dots<x-h_1,~~\forall x\in [2,\infty)$, we change variable to $y=x-h_{N_f}$ and rewrite
\beq
N=\left(\frac{\hrho^2}{\hrho-1}\right)^{N_f} d \sum_{i=2}^{\infty} \left(\hrho^i-1\right) \int_{i}^{i+1}dy~\left(\hrho^{\left|y+1-\frac{1}{N_f}\right|}-1\right)\dots\left(\hrho^{\left|y+1-\frac{N_f-1}{N_f}\right|}-1\right) \e^{-\frac{(y+h_{N_f})^2}{2\sigma^2}}
\eeq
%Note that $h_{N_f}=1/2-1/2N_f<1$ and $\hrho^{|y|-1}=0$ for $y\in[-h_{N_f}; 0)$.
Now we can split the single integration interval into $N_f$ sub-intervals in which each of the $N_f-1$ terms in the product has constant value
\beq
N=\left(\frac{\hrho^2}{\hrho-1}\right)^{N_f} d \sum_{i=2}^{\infty} \left(\hrho^i-1\right)\sum_{k=0}^{N_f-1}\int_{i+\frac{k}{N_f}}^{i+\frac{k+1}{N_f}} dy~\left(\hrho^{\left|y+1-\frac{1}{N_f}\right|}-1\right)\dots\left(\hrho^{\left|y+1-\frac{N_f-1}{N_f}\right|}-1\right) \e^{-\frac{(y+h_{N_f})^2}{2\sigma^2}}
\eeq
In the $k$-th interval $N_f-k$ terms have the value $(\hrho^i-1)$ while the remaining $k$ terms have the value $(\hrho^{i+1}-1)$. The generic term of the double sum is therefore 
\beq
\left(\hrho^i-1\right)^{N_f-k}\left(\hrho^{i+1}-1\right)^k \int_{i+\frac{k}{N_f}}^{i+\frac{k+1}{N_f}} dy~ \e^{-\frac{(y+h_{N_f})^2}{2\sigma^2}}
\eeq
changing back variable to $x=y+h_{N_f}$ we obtain
\bea
& &\left(\hrho^i-1\right)^{N_f-k}\left(\hrho^{i+1}-1\right)^k \int_{i+h_{N_f}+\frac{k}{N_f}}^{i+h_{N_f}+\frac{k+1}{N_f}} dx~ \e^{-\frac{x^2}{2\sigma^2}} \nonumber \\
&=&\left(\hrho^i-1\right)^{N_f-k}\left(\hrho^{i+1}-1\right)^k  \sqrt{\frac{\pi}{2}}\sigma 
\left[{\rm erfc}\left(\frac{i+h_{N_f}+k/N_f}{\sqrt{2}\sigma}\right)-{\rm erfc}\left(\frac{i+h_{N_f}+(k+1)/N_f}{\sqrt{2}\sigma}\right)\right]
\eea
Hence the normalization of $P^{O.T.}(L)$ for the homogenous rigid bundle is
\beq
N=d \sqrt{\frac{\pi}{2}}\sigma \left(\frac{\hrho^2}{\hrho-1}\right)^{N_f} \sum_{i=2}^{\infty} \sum_{k=0}^{N_f-1} \left(\hrho^i-1\right)^{N_f-k}\left(\hrho^{i+1}-1\right)^k  \left[{\rm erfc}\left(\frac{i+h_{N_f}+k/N_f}{\sqrt{2}\sigma}\right)-{\rm erfc}\left(\frac{i+h_{N_f}+(k+1)/N_f}{\sqrt{2}\sigma}\right)\right]
\eeq
and $P^{O.T.}(L)$ reads
\beq
P^{O.T.}(L)=\sqrt{\frac{2}{\pi}}\sqrt{\frac{\kappa_T}{k_BT}}\frac{\prod_{n=1}^{N_f}\left(\hrho^{|L/d-h^\ast_n|}-1 \right)~\e^{-\frac{(L/d)^2}{2\sigma^2}}}{\sum_{i=2}^{\infty} \sum_{k=0}^{N_f-1} \left(\hrho^i-1\right)^{N_f-k}\left(\hrho^{i+1}-1\right)^k  \left[{\rm erfc}\left(\frac{i+h_{N_f}+k/N_f}{\sqrt{2}\sigma}\right)-{\rm erfc}\left(\frac{i+h_{N_f}+(k+1)/N_f}{\sqrt{2}\sigma}\right)\right]
}
\eeq
Correspondingly, for $\<L\>^{O.T.}$ we obtain
\beq
\<L\>^{O.T.}=\sqrt{\frac{2}{\pi}}\sqrt{\frac{k_BT}{\kappa_T}}\frac{\sum_{i=2}^{\infty} \sum_{k=0}^{N_f-1} \left(\hrho^i-1\right)^{N_f-k}\left(\hrho^{i+1}-1\right)^k  \left[\e^{-\frac{\left(1+h_{N_f}+k/N_f\right)^2}{2\sigma^2}}-\e^{-\frac{\left(1+h_{N_f}+(k+1)/N_f\right)^2}{2\sigma^2}}\right]}{\sum_{i=2}^{\infty} \sum_{k=0}^{N_f-1} \left(\hrho^i-1\right)^{N_f-k}\left(\hrho^{i+1}-1\right)^k  \left[{\rm erfc}\left(\frac{i+h_{N_f}+k/N_f}{\sqrt{2}\sigma}\right)-{\rm erfc}\left(\frac{i+h_{N_f}+(k+1)/N_f}{\sqrt{2}\sigma}\right)\right]
}
\eeq

\end{document}